\documentstyle[preprint,aps,eqsecnum,epsfig]{revtex}
\tighten
\newlength{\zeichenbreit}
\newlength{\slashbreit}
\newcommand{\be}{\begin{equation}}
\newcommand{\ee}{\end{equation}}
\newcommand{\bea}{\begin{eqnarray}}
\newcommand{\eea}{\end{eqnarray}}
\newcommand{\pkt}{\; .}
\newcommand{\kma}{\; ,}
\newcommand{\intk}{\int\!\frac{d^3 k}{(2\pi)^3}\,}
\newcommand{\intt}{\int_0^t\!dt'\,}
\newcommand{\Slash}[1]{\mbox{$\settowidth{\zeichenbreit}{$#1$}
            \settowidth{\slashbreit}{$/$}\addtolength{\zeichenbreit}
            {0.9\slashbreit}#1\hspace{-0.5\zeichenbreit}\raisebox{0.1ex}
            [0ex][0ex]{/}$}}

\newcommand{\calf}{{\cal F}}
\newcommand{\calm}{{\cal M}}
\newcommand{\calv}{{\cal V}}
\newcommand{\calc}{{\cal C}}
\newcommand{\cale}{{\cal E}}
\newcommand{\bfk}{{\bf k}}
\newcommand{\eqn}[1]{(\ref{#1})}
\draft
\begin{document}
\preprint{ DO-TH 01/16, LPTHE-01-47, LA-UR-01-5699}
\title{\bf Out of Equilibrium Dynamics of Supersymmetry 
at High Energy Density}
\author{{\bf J. Baacke}$^{(a)}$, 
{\bf D. Cormier}$^{(b)}$, {\bf H. J. de Vega}$^{(c)}$, and 
{\bf K. Heitmann}$^{(d)}$}
\address
{(a) Institute for Physics, University of Dortmund, D-44221 Dortmund,
Germany\\
(b) Centre for Theoretical Physics, Sussex University, Falmer,
Brighton, BN1 9QH, England\\
(c) Laboratoire de Physique Th\'eorique et Hautes Energies,
Universit\'e Pierre et Marie Curie (Paris VI) et Denis Diderot (Paris VII),
Tour 16, 1er. \'etage, 4, Place Jussieu, 75252 Paris cedex 05, France\\ 
(d) T-8, Theoretical Division, Los Alamos National Laboratory,
Los Alamos, New Mexico 87545, USA}
\date{October, 2001}
\maketitle
\begin{abstract}
  We investigate the out of equilibrium dynamics of global chiral
  supersymmetry at finite energy density.  We concentrate on two
  specific models.  The first is the massive Wess-Zumino model which
  we study in a selfconsistent one-loop approximation.  We find that
  for energy densities above a certain threshold, the fields are
  driven dynamically to a point in field space at which the fermionic
  component of the superfield is massless.  The state, however is
  found to be unstable, indicating a breakdown of the one-loop
  approximation.  To investigate further, we consider an $O(N)$
  massive chiral model which is solved exactly in the large $N$ limit.
  For sufficiently high energy densities, we find that for late times
  the fields reach a nonperturbative minimum of the effective
  potential degenerate with the perturbative minimum. This minimum is
  a true attractor for $O(N)$ invariant states at high energy
  densities, and this provides a mechanism for determining which of
  the otherwise degenerate vacua is chosen by the dynamics. The final
  state for large energy density is a cloud of massless particles
  (both bosons and fermions) around this new nonperturbative
  supersymmetric minimum. By introducing boson masses which softly
  break the supersymmetry, we demonstrate a see-saw mechanism for
  generating small fermion masses.  We discuss some of the
  cosmological implications of our results.
\end{abstract}


\section{Introduction}
Supersymmetry\cite{susy}, by nature of its Grassman transformation
parameter, behaves differently from ordinary symmetries under the
effects of finite temperature, as was first recognized over 20 years
ago by Das and Kaku\cite{daskaku} and has been studied since by a
number of
authors\cite{GGS,fujikawa,hove,fuchs,boyan,chia,GS,das,LR,BO,DL}.  In
particular, it was shown that unbroken supersymmetry in equilibrium at
zero temperature becomes broken at finite
temperatures\cite{boyan,chia}.  As usual, such breaking of the
continuous symmetry is necessarily signaled by the appearance of a
massless particle, the Goldstone.  Only in this case the Goldstone is
a fermion rather than a boson, due to the fact that the symmetry
transformation parameter is a Grassman variable; the residual flat
direction in field space must correspond to a fermionic degree of
freedom.  The Goldstone fermion is referred to as the Goldstino.

The fundamental reason why supersymmetry becomes broken at finite
temperature is easy to understand.  At finite temperature, fermions
and bosons obey different statistics.  This means that while the
Lagrangian may admit a supersymmetry between boson and fermion
components, the way in which these components are populated at finite
temperature breaks any such symmetry.  In equilibrium, the only state
which transforms supersymmetrically is the zero temperature
supersymmetric ground state.

These issues are relevant for a number of reasons.  We see no
supersymmetric partners to the fermions nor to the bosons in nature.
Hence, if we are to assume that supersymmetry is fundamental, then it
must be broken. Supersymmetry breaking is an appealing possibility
which raises the question of the nature and identification of the
resulting Goldstino.

The possible breaking of supersymmetry is also relevant to the early
universe.  In particular, many popular inflation models are based on
supersymmetry in some form\cite{lythriotto}.  However, inflation
occurs very far from thermal equilibrium and while supersymmetric
models are understood at zero and finite temperature, the
non-equilibrium dynamics of such models have yet to be properly
studied.  And even when supersymmetric models are considered in the
context of inflation, it is common practice to discard some of the
degrees of freedom as irrelevant.  However, if the results in thermal
equilibrium are to be a guide, this may be a dangerous thing to do.
Processes such as the breaking of supersymmetry and the consequent
appearance of massless degrees of freedom might also occur out of
equilibrium as the available energy is distributed differently among
fermions and bosons.  This may be particularly relevant for the
process of reheating after inflation, for which light fermions may
play an important
role\cite{Baacke:98c,pbgreene,jgarcia,stsujikawa,pbgreen2}.

We examine these issues by means of explicit numerical solutions of
supersymmetric toy models allowed to evolve far from equilibrium.
First results describing the dynamics of $O(N)$ chiral supersymmetry
at finite energy were published recently \cite{BCdVH}.  Although the
primary applications of interest are cosmological, we simplify the
analysis by restricting ourselves to Minkowski spacetime.

We begin our analysis with a study of the massive Wess-Zumino model in
a self-consistent one-loop approximation.  We find two distinct
dynamical regimes.  At sufficiently low energy density, the system
remains near one of the two supersymmetric vacua of the model.  The
non-zero energy is taken up in the oscillation of the field zero modes
about their perturbative vacuum values.  Some particles are produced,
but there is no obvious massless fermion in the spectrum over the
lifetime of the numerical simulations.  These states are continuously
connected to the zero temperature vacuum states in the sense that as
one reduces the initial energy density of the system toward zero, the
system approaches the zero temperature vacuum without any changes in
the qualitative behavior of the system.

The second regime is quite different.  When the system exceeds a
certain critical energy density such that the fields can sufficiently
sample the region separating the two perturbative minima, the system
becomes driven to the point halfway between the two minima, i.e., to
the local maximum of the effective potential.  What is special about
this point is that the mass of the fermion field precisely vanishes.

There are a couple things we learn from this behavior.  First, this
provides another instance of the dangers of relying solely on
equilibrium constructs such as the effective potential when dealing
with systems far from equilibrium\cite{neqvseffpot,LosAl}.  The
perturbative effective potential alone would never cause one to expect
that the system would approach such a point.  Second, with sufficient
energy to reach the relevant point in field space, the system will
invariably approach a state for which the fermion field is massless.
Numerical tests of the system indicate that this behavior is obtained
for arbitrarily high energy densities.

However, we also find that this state has an instability to this order
in perturbation theory as it includes a scalar field with negative
mass squared, corresponding to a so-called spinodal instability.  This
is expected from being at the local maximum of the effective
potential.  This is a signal of the breakdown of perturbation theory
and an indication that the state which eventually forms is
non-perturbative in nature.  This unfortunately limits our study in
the one component model to the early time behavior before the
instability becomes important.

In order to proceed beyond the perturbative Wess-Zumino model, we
introduce a model with an internal $O(N)$ symmetry, which can be
solved exactly, even into the non-perturbative regime, in the limit $N
\to \infty$.  While an $O(\infty)$ field theory may not be realistic,
it provides a useful self-consistent toy model for the study of
possible characteristics of realistic theories -- particularly those
with continuous symmetries -- and may also be considered as a first
approximation to systems with moderate values of $N$.  Such techniques
have a long history in field theory\cite{largeN} in general as well as
in supersymmetric theories\cite{largeNsusy}.  In the case of purely
scalar models, large $N$ studies provide concrete examples of the
non-perturbative symmetry breaking processes leading to Goldstone
bosons\cite{LosAl,largeNgoldstone} and to a dynamical formation of a
flat potential related to the Maxwell construction of equilibrium
thermodynamics\cite{dynmaxwell}.

Our results again indicate two regimes for states invariant under the
$O(N)$ symmetry.  As with the one-loop Wess-Zumino model, there is a
continuous spectrum of states at sufficiently low energy densities
which are connected to the perturbative $O(N)$ symmetric vacua. For
sufficiently high energy densities, the system again is driven to a
point in field space for which the fermion modes are massless. This
point, $ \phi = -m/\lambda $ which acts as an attractor to the
dynamics is a nonperturbative minimum of the effective potential
degenerate with the perturbative minimum. The final state obtained by
real time evolution for large energy density is a cloud of massless
particles (both bosons and fermions) around this new nonperturbative
minimum.

There are several interesting facts about this system that deserve
mention.  First, we see that the dynamics effectively chooses the
vacuum for the system during the high energy density phase.  In a
cosmological context, once such a vacuum is chosen, the universe would
stay in that vacuum indefinitely.  We therefore see a mechanism for
choosing one vacuum over another in spite of the fact that they are
degenerate in energy.  Second, we find that all particles are massless
for arbitrarily high energy density.  This is to be contrasted with
the general, non-supersymmetric case for which continuous symmetries
are always restored at high enough energy density and particles become
massive.  The study of the spontaneously broken $O(N)$ scalar theory
provides a good example of this more usual behavior\cite{dynmaxwell}.
In the present case, however, the supersymmetry acts to protect the
kind of Goldstone phase we find with all particles, bosons and
fermions, being massless.

As a final example, we examine what occurs when the supersymmetry of
the $O(N)$ model is softly broken by the introduction of additional
small scalar mass terms.  The result is that while there is still a
set of massless bosons in the high energy density phase, their
superpartner bosons and fermions gain masses.  While the bosons gain a
mass proportional to the soft breaking mass scale, there is a see-saw
mechanism for the fermions which gives them a mass proportional to the
square of the soft breaking mass scale divided by the overall scale of
supersymmetry, providing a very natural mechanism for producing very
light fermions which could be relevant to neutrino mass generation.

We continue in the next section with the introduction of the
Wess-Zumino model and provide the renormalized one-loop equations of
motion appropriate to an out of equilibrium study.  This is followed
by our numerical results for the model, showing the two distinct
regimes at low and high energies.  In Section \ref{sec3}, we introduce
the $O(N)$ model and again provide our numerical results.  In Section
\ref{sec4}, we include soft supersymmetry breaking into the $O(N)$
Lagrangian and show how the various fields gain non-zero masses.  Our
conclusions are provided in Section \ref{sec5}.  We also present two
appendices with details of the renormalization procedures used in the
non-equilibrium formalism.

\section{The Wess-Zumino model to one loop order} \label{sec2}
\subsection{model and equations of motion}

We consider the supersymmetric Wess-Zumino model\cite{Wess:1974}.  It
is based on a single chiral super-multiplet $S$ with
superpotential\cite{susy}
\begin{equation}\label{WZlag}
W(S) = \frac12 m S \cdot S + \frac13 \lambda S \cdot S \cdot S \; ,
\end{equation}
This can be broken down into component fields via
$S = \left({\cal A},{\cal B};\psi;F,G\right)$, 
where ${\cal A}$ is a scalar, ${\cal B}$ is a pseudo-scalar, 
$\psi$ is a Majorana fermion, and $F$ and $G$ are scalar and 
pseudo-scalar auxiliary fields respectively.  After eliminating 
the auxiliary fields, the Lagrangian is given by
\begin{eqnarray}
{\cal L}&=& \frac12 \partial_\mu{\cal A}\partial^\mu{\cal A} 
+\frac12 \partial_\mu{\cal B}\partial^\mu{\cal B}
-\frac12 m^2 \left({\cal A}^2 + {\cal B}^2\right)
-\frac{1}{2}\lambda m {\cal A}\left({\cal A}^2 + {\cal B}^2\right)
-\frac18 \lambda^2 \left({\cal A}^2 + {\cal B}^2\right)^2 \nonumber\\
&&+\frac{i}{2}\bar\psi\Slash\partial\psi-\frac 1 2 m\bar\psi\psi
-\frac{\lambda}{2}{\cal A} \bar\psi\psi
-\frac{i\lambda}{2}{\cal B}\bar\psi\gamma_5\psi \pkt
\end{eqnarray}
In order to follow the dynamics, it is convenient to break up the
scalar field ${\cal A}$ into its expectation value and small fluctuations
about that value:
\begin{equation}
{\cal A} = \phi + A \; , \, \, \phi = \langle {\cal A} \rangle \; , 
\, \, \langle A \rangle = 0 \; .
\end{equation}
For convenience, we take ${\cal B}$ to have zero expectation
value.  We write 
\begin{equation}
{\cal B} = B \; , \, \, \langle B \rangle = 0 \; .
\end{equation}
Likewise, we treat $\psi$ as a pure fluctuation with 
$\langle \psi \rangle = 0$.
We can then expand the Lagrangian in orders of the fluctuations, $A$,
$B$, and $\psi$. 

In zeroth order we find
\begin{equation}
{\cal L}^{(0)}=\frac 1 2\dot\phi^2-V(\phi)\kma
\end{equation} 
with the classical potential
\begin{equation}
V(\phi)=\frac{1}{2}\phi^2\left(m+\frac{\lambda}{2}\phi 
\right)^2\kma
\end{equation}
as sketched in Fig.~1.

\noindent
\parbox{14.8cm}{
%\parbox[t]{8cm}
{\begin{center}
\mbox{\epsfxsize=7.5cm\epsfbox{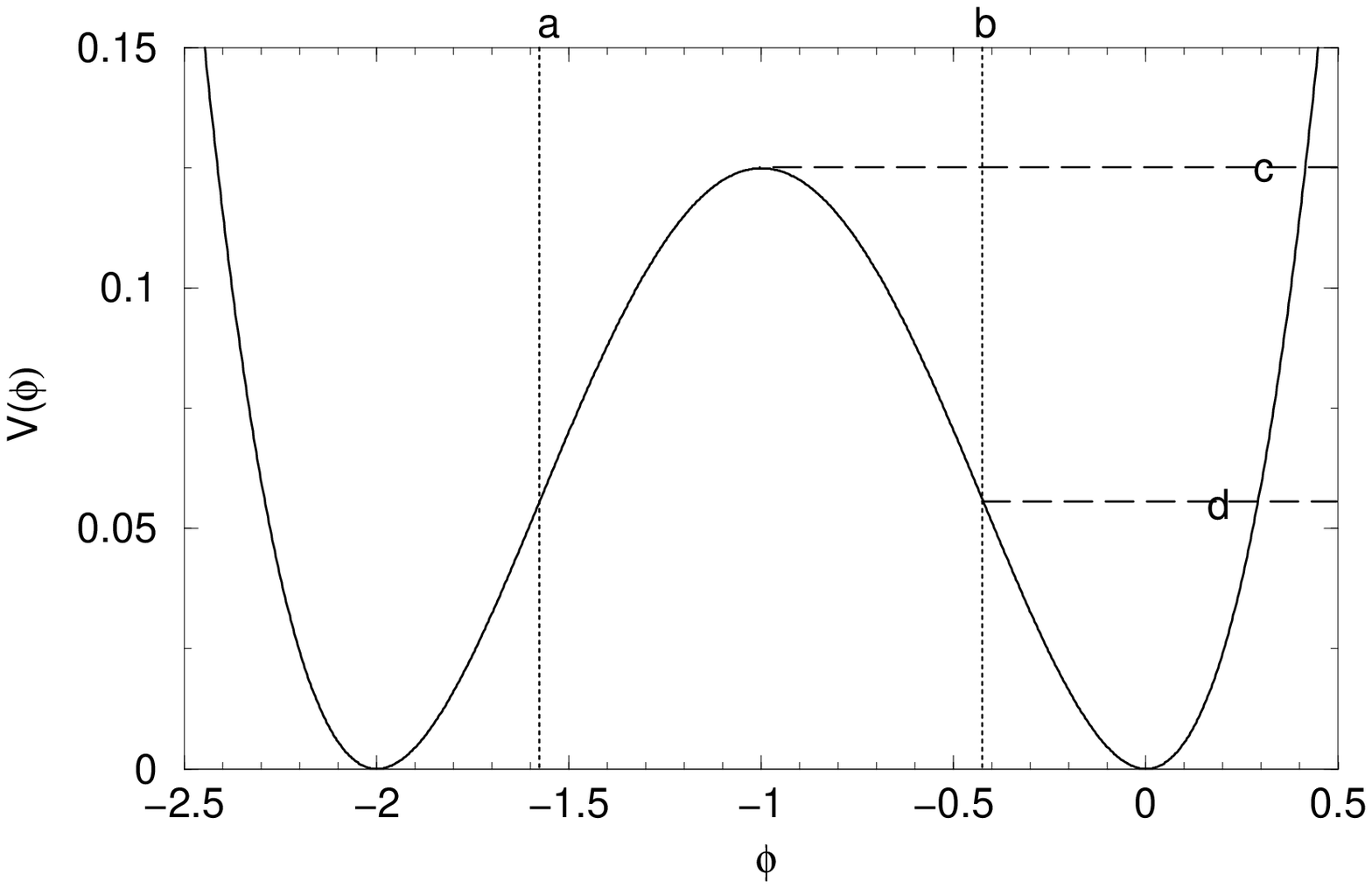}}
\end{center}}
\hspace{.0cm} {{\small FIG. 1: The classical scalar potential with
$\lambda=m=1.0$.  The dotted line represents the spinodal line above
which there is sufficient energy to enter the negative curvature
portion of the potential, $a=-m(1+1/\sqrt{3})/\lambda$,
$b=-m(1-1/\sqrt{3})/\lambda$. The upper dashed line indicates the
initial value for which the energy is high enough for $\phi$ to pass
the maximum, $c=m^4/8\lambda$, the lower dashed line marks the upper
limit for the initial value for $\phi$, which leads to a stable
configuration, $d=m^4/18\lambda$. Initial values in the region above
d lead to unstable configurations.}}}

\vspace{.5cm}

This yields the classical part of the equation of motion.
The first order in the fluctuation vanishes because the expectation
value of the fluctuations is zero. The second order expression reads
\begin{eqnarray}
{\cal L}^{(2)}&=&
\frac 1 2 \partial_\mu A\partial^\mu A
+\frac 1 2 \partial_\mu B\partial^\mu B
-\frac{m^2}{2}\left( A^2+ B^2\right)\nonumber\\
&&-\frac{1}{2}m\lambda\phi\left(3 A^2+ B^2\right)
-\frac{\lambda^2}{4}\phi^2\left(3 A^2+B^2\right)\nonumber\\
&&+\frac i 2\bar\psi\Slash\partial\psi-\frac 1 2 m\bar\psi\psi-\frac\lambda 2
\phi\bar\psi \psi \pkt
\end{eqnarray}

We can derive the equation of motion for
the classical field and  for the fluctuations from these Lagrangians.
We find for the classical field in one-loop approximation:
\begin{equation}
\ddot\phi+V'(\phi)
+\frac{\lambda}{2}\left(m+\lambda \phi\right)\left[3\langle A^2\rangle
+\langle B^2\rangle\right]
+\frac{\lambda}{2} \langle\bar\psi\psi\rangle=0\kma
\label{oneloopeom}
\end{equation}
while for the fluctuations we find
\begin{eqnarray} 
\ddot A+\left(-{\nabla}^2+m^2+\frac32 \lambda^2\phi^2
+3 m\lambda\phi\right) A&=&0\kma\\
\ddot B+\left(-{\nabla}^2+m^2+\frac12 \lambda^2\phi^2
+ m\lambda\phi\right) B&=&0\kma\\
\left(i\Slash\partial-m- \lambda\phi\right)\psi&=&0\pkt
\end{eqnarray}

It is convenient to collect the mass term and the terms depending on
$\phi(t)$ into  time-dependent masses
\begin{eqnarray}
\label{ma}
\calm_A^2(t)&=&m^2+3 m\lambda\phi(t)+\frac 3 2\lambda^2\phi^2(t)\kma\\
\label{mb}
\calm_B^2(t)&=&m^2+m\lambda\phi(t)+\frac{\lambda^2}{2}\phi^2(t) \kma \\
\label{mpsi}
\calm_\psi(t)&=&m+\lambda \phi(t)
\pkt \label{psimass}
\end{eqnarray}
We note that the supersymmetry relation
\begin{equation}
\calm^2_A + \calm^2_B - 2 \calm^2_\psi = 0 \; ,
\label{susysumrule}
\end{equation} 
is satisfied by the time-dependent masses for all times.
We further note that in the region 
\begin{equation}
-1-\frac{1}{\sqrt{3}} < \lambda \phi/m < -1+\frac{1}{\sqrt{3}} \; ,
\label{spinodalregion}
\end{equation}
$\calm_A^2$ is negative, leading to instabilities.
$\calm_B^2(t)$ on the other hand is everywhere positive definite.

We provide integral expressions for the expectation values appearing
in Eq.~(\ref{oneloopeom}) below.  Note that if the fermion mass
$\calm_\psi(t)$ vanishes, i.e. $\phi(t)$ is equal $-m/\lambda$, the
contribution of the bosonic fields $ A$ and $ B$ cancels in
(\ref{oneloopeom}).  In this case, only the fermionic fluctuations
influence the equation of motion and the behavior of the classical
field.

We expand the scalar fields in terms of mode functions $f_A$ and $f_B$,
and the fermion field in terms of the spinor solutions of the Dirac
equation as in \cite{Baacke:98c,Baacke:97a}
\begin{eqnarray}
 A(t,{\bf x})&=&\intk\frac{1}{2\omega_{ A 0}}\left[
c_{A,{\bf k}}f_A(t)+c^\dagger_{A,{\bf k}}f_A^*(t)\right]
e^{i{\bf k\cdot x}}\kma \label{Aexp} \\
 B(t,{\bf x})&=&\intk\frac{1}{2\omega_{ B 0}}\left[
c_{B,{\bf k}}f_B(t)+c^\dagger_{B,{\bf k}}f^*_B(t)\right]
e^{i{\bf k\cdot x}}\kma\\
\psi(t,{\bf x})&=&\intk\frac{1}{2\omega_{\psi 0}}\left[
b_{{\bf k},s}U_{{\bf k},s}(t)+b^\dagger_{{-\bf k},s}V_{{-\bf k,s}}
(t)\right]e^{i{\bf k\cdot x}}\kma
\end{eqnarray}
with the usual (anti-)commutation relations for the 
time independent annihilation and creation operators
\begin{eqnarray}
[{c_{j,{\bf k}},c^\dagger_{j,{\bf k'}}}]&=&2\omega_{j0}(2\pi)^3
\delta^3({\bf k}-{\bf k'})  \, \, , \; j=A,B \kma \\
\{b_{{\bf k},s},b^\dagger_{{\bf k'},s}\}&=&2\omega_{\psi 0}(2\pi)^3
\delta^3({\bf k}-{\bf k'})\delta_{ss'}\pkt \label{fermcomrel}
\end{eqnarray}
The frequencies in Eqs.~(\ref{Aexp})--(\ref{fermcomrel}) 
are defined as
\begin{equation}
\omega^2_{j0} = k^2 + m^2_{j0} \, \, , \; j = A,B,\psi \; ,
\end{equation} 
with $m_{j0}=\calm_j(0)$. 
%We occasionally use the definition
%\be
%\omega_j^2(t)= k^2 +\calm_j^2(t)
%\pkt
%\ee

The equations of motion for the bosonic mode functions are given by
\begin{equation}
\label{modesc}
\ddot f_j(t)+\left[k^2+\calm_j^2(t)\right]f_j(t)=0\kma
\end{equation}
with $j= A, B$.
The initial conditions for the fields are chosen as
\begin{equation}\label{init}
f_j(0)=1\kma~~~\dot f_j(0)=-i\omega_{j0}\pkt
\end{equation}

For the fermions, we define mode functions $f_\psi(k,t)$ and
$g_\psi(k,t)$ through the relations
\bea
U_s(\bfk,t)&=&N_0\left[i\partial_t+ 
{\cal H}_\bfk(t)\right]f_\psi(k,t)
{\chi_s \choose 0}
\kma
\\
V_s(\bfk,t)&=&N_0\left[i\partial_t+ 
{\cal H}_{-\bfk}(t)\right]g_\psi(k,t)
{0 \choose \chi_s}
\kma
\eea
where the $\chi_s$ with $s=\pm 1$ are helicity eigenstates with eigenvalue
$s$ and ${\cal H}_\bfk(t)$ is defined as
\be
{\cal H}_\bfk(t) = \gamma_0 \mbox{\boldmath $\gamma$\unboldmath}
\cdot {\bf k} + \gamma_0 \calm_\psi 
\pkt \ee

The mode functions $f_\psi(k,t)$ obey the second order differential equation
\be\label{fsec}
\left[
\frac{d^2}{dt^2}-i\dot\calm_\psi(t)+ k^2+\calm_\psi^2(t)
\right]f_\psi(k,t)=0\kma 
\ee
while $g_\psi(k,t) = f^*_\psi(k,t)$.
The initial conditions are
\be\label{fginit}
f_\psi(k,0)=1,~~~ \dot f_\psi(k,0)=-i\omega_{\psi 0}\pkt
\ee

The integrals appearing in the equation of motion for $\phi$, 
Eq. (\ref{oneloopeom}), are:
\begin{eqnarray}
\langle A^2 \rangle(t) &=& \intk \frac{|f_A(t)|^2}{2\omega_{ A 0}} 
\kma \label{afluct} \\
\langle B^2 \rangle(t) &=& \intk \frac{|f_B(t)|^2}{2\omega_{ B 0}} 
\kma \label{bfluct} \\
\langle \overline{\psi} \psi \rangle(t) &=&- \intk 
\frac{1}{\omega_{\psi 0}}\left[2\omega_{\psi0}
-\frac{2k^2}{\omega_{\psi 0}+m_{\psi 0}}|f_\psi(t)|^2\right] 
\pkt \label{psifluct}
\end{eqnarray}

The energy density can be calculated as the trace over the
Hamiltonian.  With the results for the scalar fields in
\cite{Baacke:97a} and the fermion fields in \cite{Baacke:98c} we can
express the energy density in terms of the mode functions in the
following way
\begin{eqnarray}
{\cal E}&=&\frac 1 2 \dot\phi+V[\phi(t)]\nonumber\\
&&+\intk\frac{1}{2\omega_{ A 0}}\left[\frac12 |\dot f_A|^2
+\frac12 \left(k^2+\calm_A^2(t)\right)|f_A|^2\right]\nonumber\\
&&+\intk\frac{1}{2\omega_{ B 0}}\left[\frac12 |\dot f_B|^2
+\frac12 \left(k^2+\calm_B^2(t)\right)|f_B|^2\right]\nonumber\\
&&+\intk\frac{1}{2\omega_{\psi 0}}\left[
i\left(\omega_{\psi 0}-m_{\psi 0}\right)\left(f_\psi\dot f_\psi^*
-\dot f_\psi f_\psi^*\right)-2\omega_{\psi 0}\calm_\psi(t)\right]
\pkt \label{efluct}
\end{eqnarray}
It is easy to verify that the time derivative of the energy vanishes
if the equation of motion for $\phi$ and those of the fluctuation
modes are satisfied.

Both the equation of motion for the classical field and for the energy
density are divergent. Therefore we have to consider their
renormalization.  Within a computational scheme based on
\cite{Baacke:90} and extended in \cite{Baacke:97a} for non-equilibrium
dynamics it is possible to make a clean separation between divergent
and finite parts, allowing us to directly compute the finite parts of
the integrals appearing in eqs.(\ref{afluct})-(\ref{psifluct}).
Details of the scheme, developed for various physical models in
refs.~\cite{Baacke:98c,Baacke:97a,Baacke:97b,Heit,Baacke:97c,Baacke:99,Baacke:98b,Baacke:00},
are presented in the Appendix A.

The structure of the renormalized equations is found to be analogous
to that of the unrenormalized ones, so we do not display them here.
The divergent expectation values appearing in
Eqs.~(\ref{afluct})--(\ref{efluct}) are replaced by finite, subtracted
expressions.  The resulting divergent counter terms are the same as
the standard ones obtained in equilibrium. It is worth mentioning that
supersymmetry leads to the expected cancellations of quadratic and
quartic divergences leaving only the relatively well behaved
logarithmic divergences -- the primary advantage of supersymmetry
models.

\subsection{Results and interpretation}

We find two distinct phases possible in the dynamics of the
Wess-Zumino model.  We examine each in turn.

\noindent
\begin{center}
\parbox{14.8cm}{
\parbox[t]{7cm}
{\begin{center}
\mbox{\epsfxsize=6.5cm\epsfbox{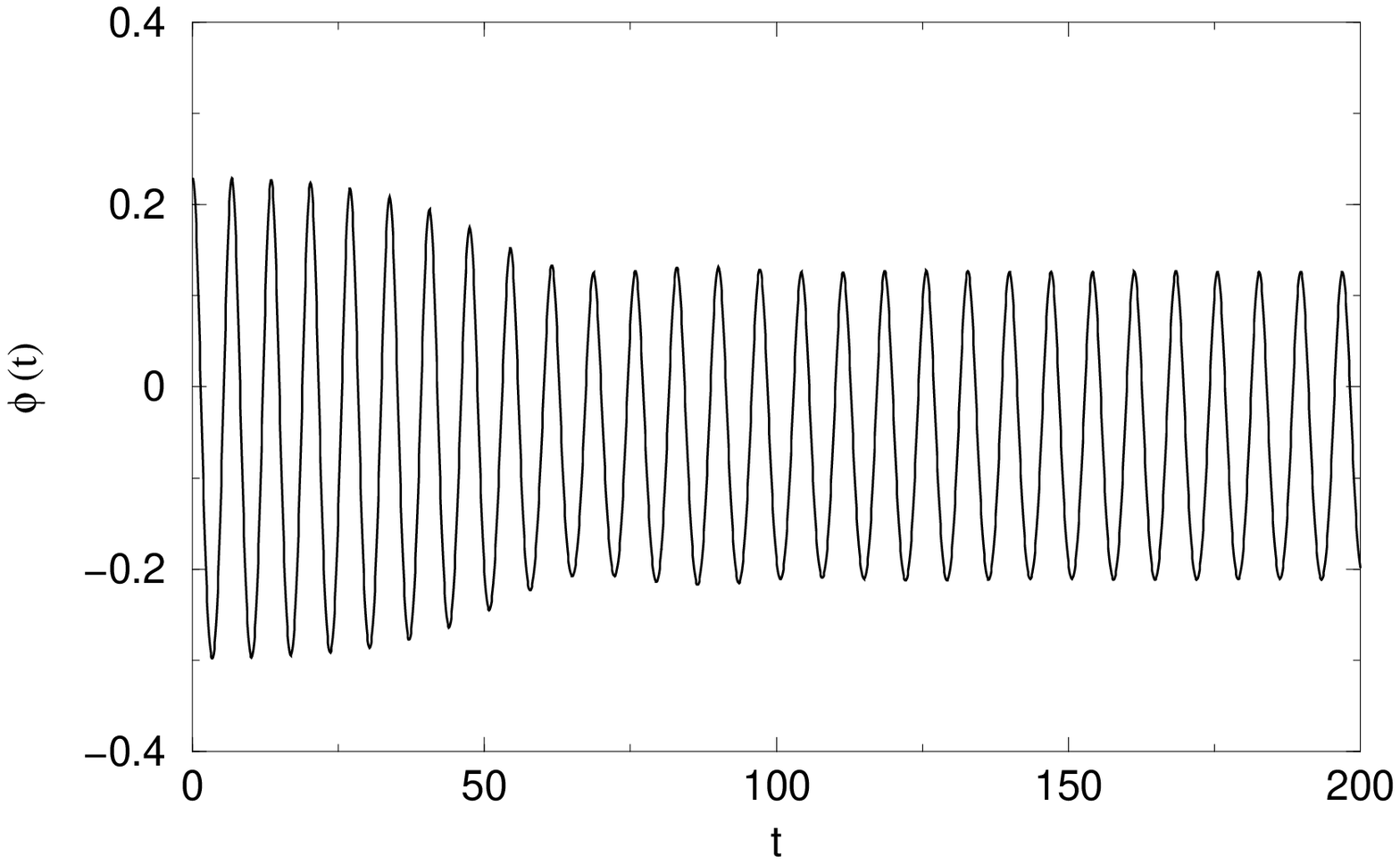}}
\end{center}}
\hspace{.0cm}\parbox[t]{7cm}
{\begin{center}
\mbox{\epsfxsize=6.5cm\epsfbox{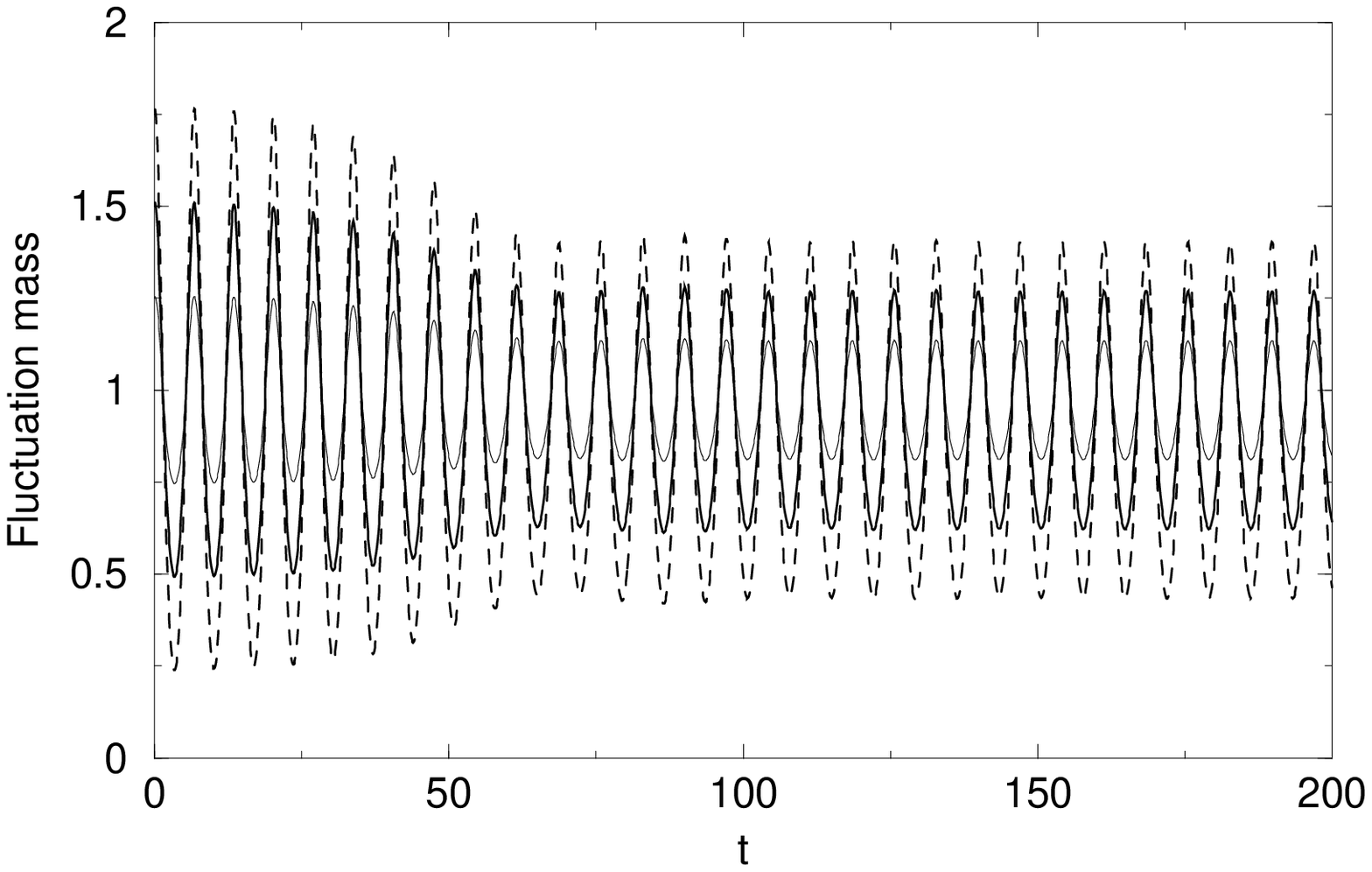}}
\end{center}}
\parbox[t]{7cm}
{{\small
FIG. 2: Zero mode evolution showing the
low energy density phase for $m=\lambda=1.0$ and $\phi(0)=0.23$.
}}
\hspace{0.5cm}\parbox[t]{7cm}
{{\small FIG. 3: The effective squared masses; 
solid line: $\calm_\psi^2(t)$, dotted line: $\calm_A^2(t)$, 
dashed line: $\calm_B^2(t)$; the parameters 
as in Fig.~2.
}
}}
\end{center}

\vspace{0.5cm}

At low energy density the model behaves much as one would expect from
the scalar potential shown in Fig.~1.  The zero mode $\phi$ oscillates
about one of the two vacua, see Fig.~2.  Particle production is
minimal.  As a result of the deviation of the zero mode from the
vacuum, the masses of the various quanta each oscillate about their
supersymmetric values as shown in Fig.~3.  Note that while there is a
small mass splitting between the masses, the overall supersymmetry sum
rule, (\ref{susysumrule}) remains satisfied.  As the overall energy
density is reduced to zero, these mass splittings vanish, indicating
that one reaches the true supersymmetric vacuum.

These numerical results indicate that the system goes to a limiting
cycle. Namely, the zero mode keeps oscillating forever. This is in
contrast to the $\Phi^4$ model in either the large $N$ limit or in the
Hartree approximation for which the zero mode always has a constant
infinite time limit\cite{dynmaxwell,asinto}.

The behavior of the system is very different, however, if the initial
energy density is large enough.  In particular, we find that whenever
the initial energy density is sufficiently large, there is an
attractor solution which causes the system to fall into a state
signaled dramatically by the vanishing of the fermion mass.
Examination of the classical potential for $\phi$ (Fig.~1) and the
expression for the effective mass of the fermion $\psi$
(\ref{psimass}) reveals that the only point for which the fermion is
massless corresponds precisely to the local maximum of the classical
potential for $\phi$!

\noindent
\begin{center}
\parbox{15cm}{
\parbox[t]{7cm}
{\begin{center}
\mbox{\epsfxsize=7cm\epsfbox{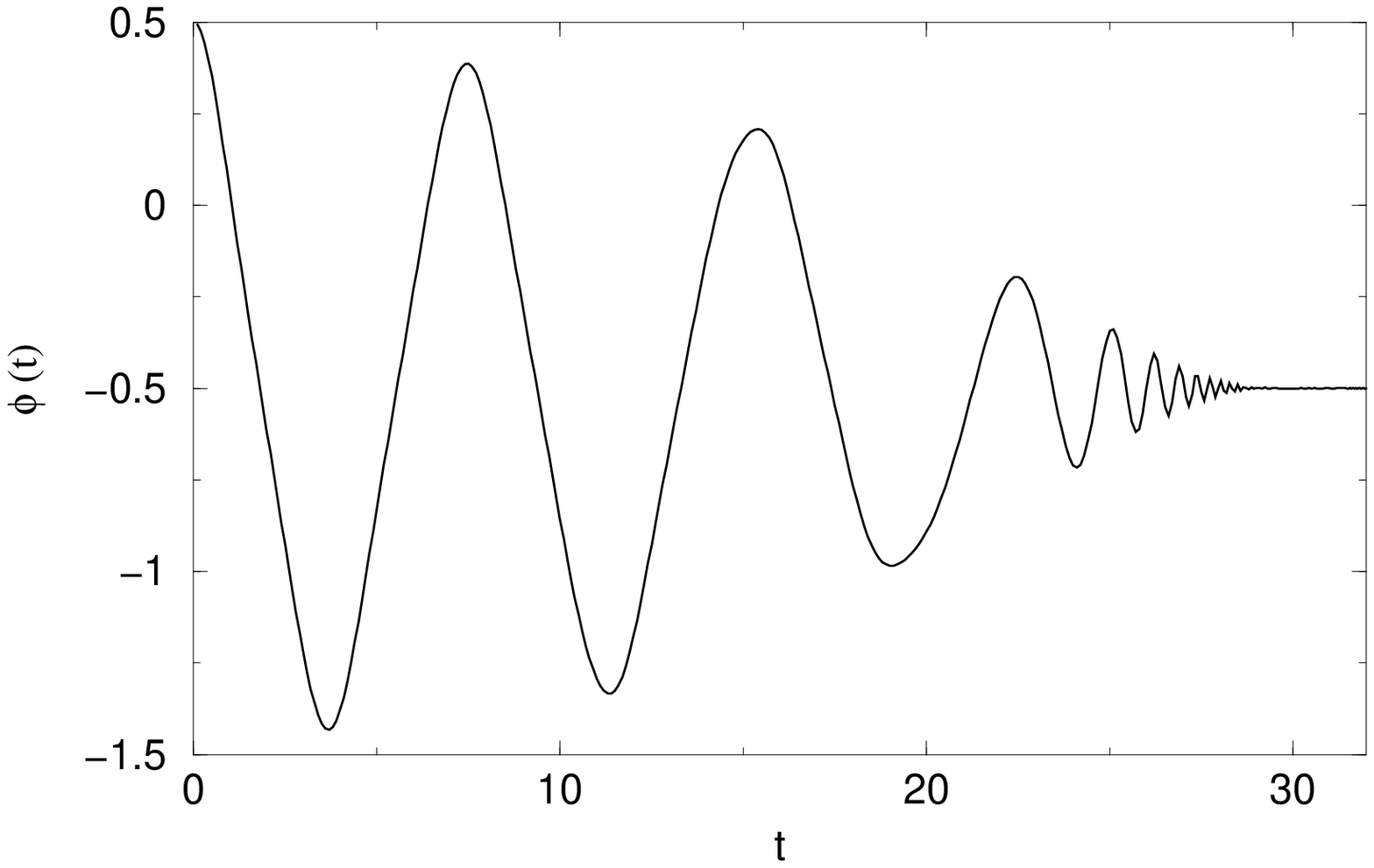}}
\end{center}}
\hspace{.0cm}\parbox[t]{7cm}
{\begin{center}
\mbox{\epsfxsize=7cm\epsfbox{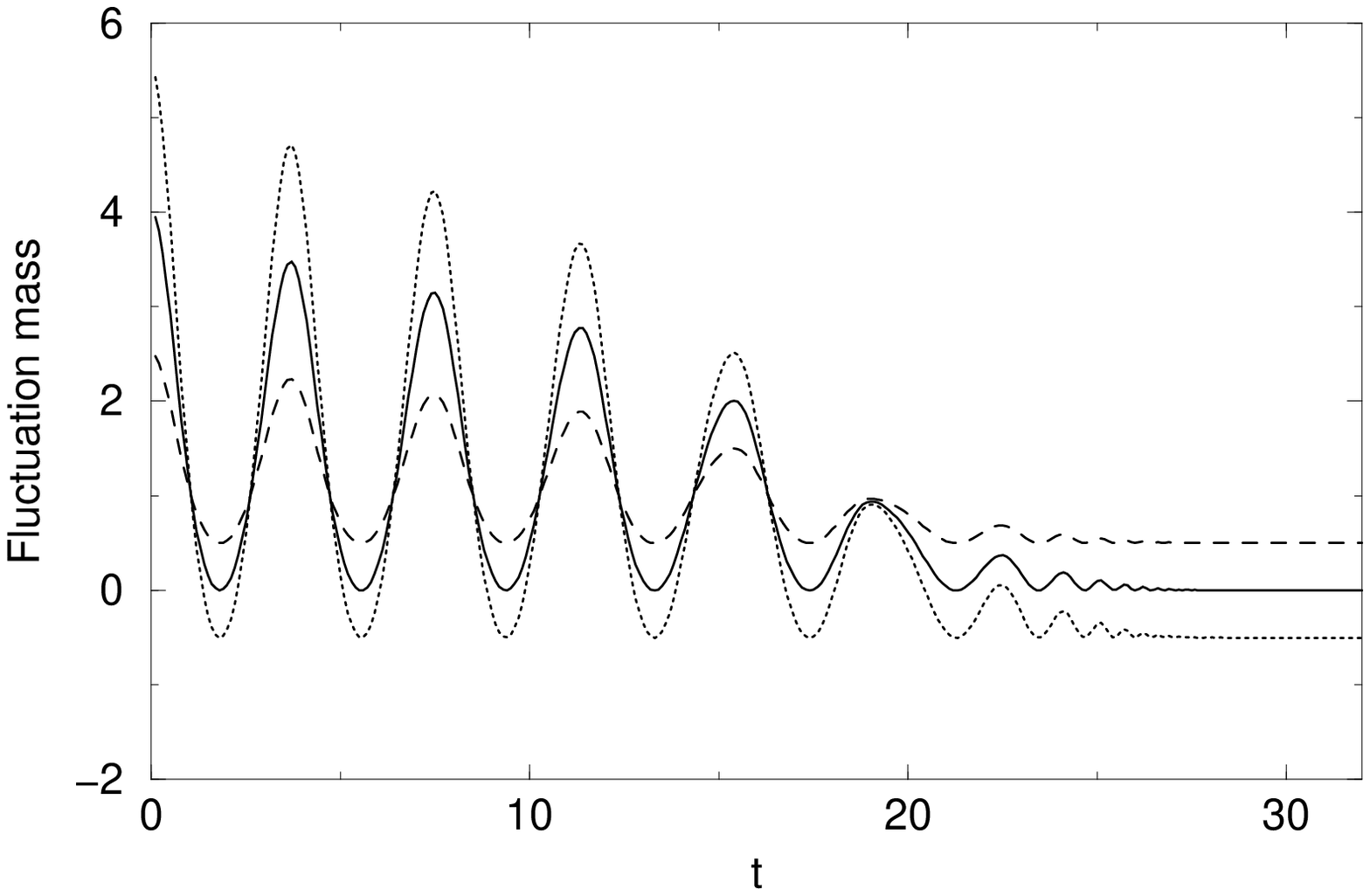}}
\end{center}}
\parbox[t]{7cm}
{{\small
FIG. 4: Zero mode evolution showing the
high energy density phase for $m=1.0$, $\lambda=2.0$ and $\phi(0)=0.5$.
}}
\hspace{0.5cm}\parbox[t]{7cm}
{{\small FIG. 5: The effective squared masses; 
solid line: $\calm_\psi^2(t)$, dotted line: $\calm_A^2(t)$, 
dashed line: $\calm_B^2(t)$; parameters 
as in Fig.~4. Note that $\calm_\psi \to 0$ while
$\calm_A^2$ becomes negative.
}
}}
\end{center}

\vspace{0.5cm}

The results of a sample evolution is plotted in Fig.~4 with the
effective masses of the fields given in Fig.~5.  We find, indeed, that
$\phi$ goes to the point $-m/\lambda$ with the mass of the fermion
approaching zero.  We also see that the evolution starts to blow up
toward the end of the simulations as demonstrated in Fig.~6 (and this
unstable behavior continues if allowed to evolve further).  This is
due simply to the fact that $\phi$ is sitting on top of its potential,
leading to a severe instability in its fluctuations $\langle A^2
\rangle$ due to the effective negative mass squared.  This is also the
signal of the breakdown of the one loop approximation.  In fact, it is
expected that perturbation theory as a whole will break down at this
point due to the process known as spinodal
decomposition\cite{spindecomp}, the result being the formation of a
fully non-perturbative and inhomogeneous state.

\noindent
\begin{center}
\parbox{15cm}{
\parbox[t]{7cm}
{\begin{center}
\mbox{\epsfxsize=7cm\epsfbox{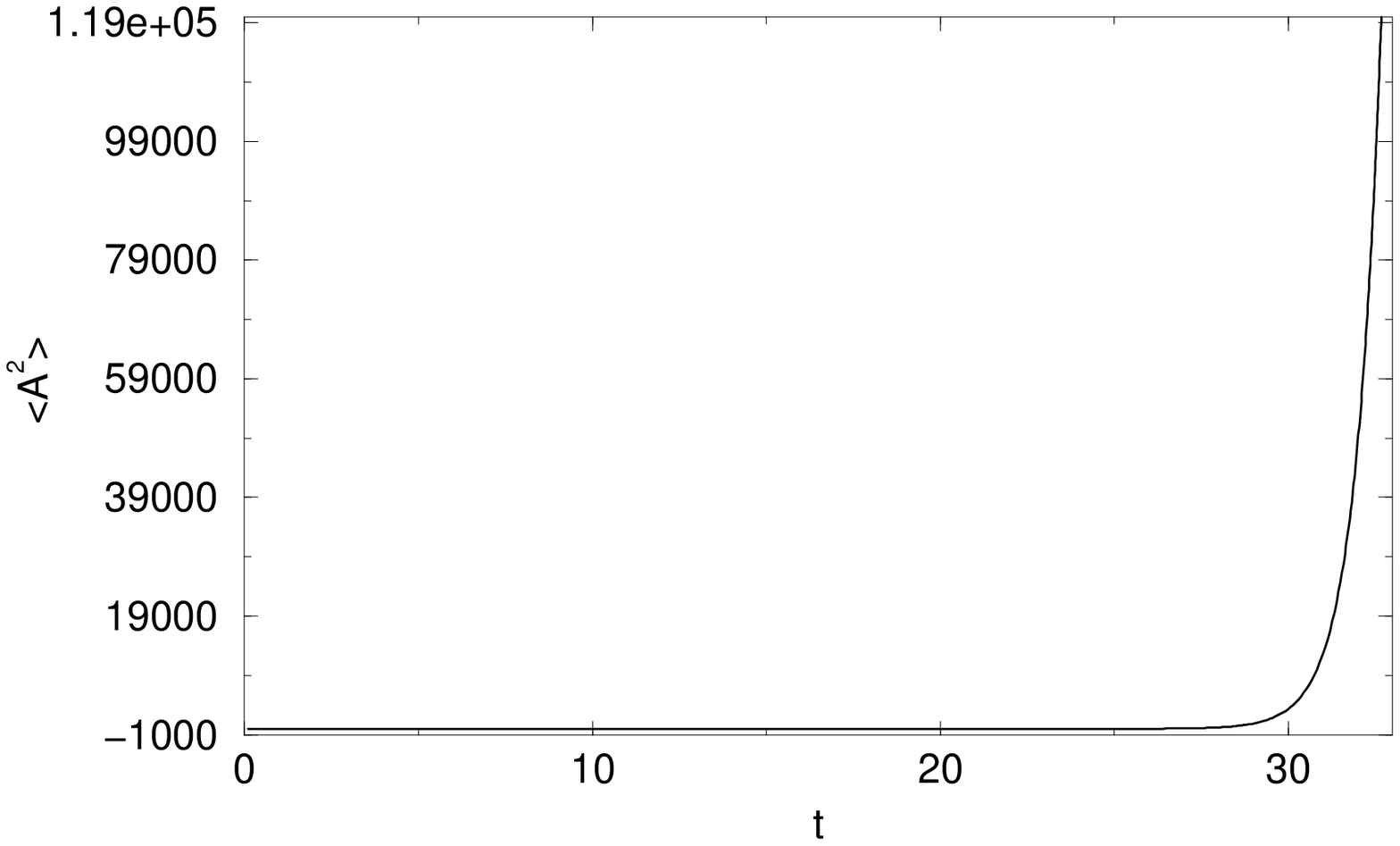}}
\end{center}}
\hspace{.0cm}\parbox[t]{7cm}
{\begin{center}
\mbox{\epsfxsize=7cm\epsfbox{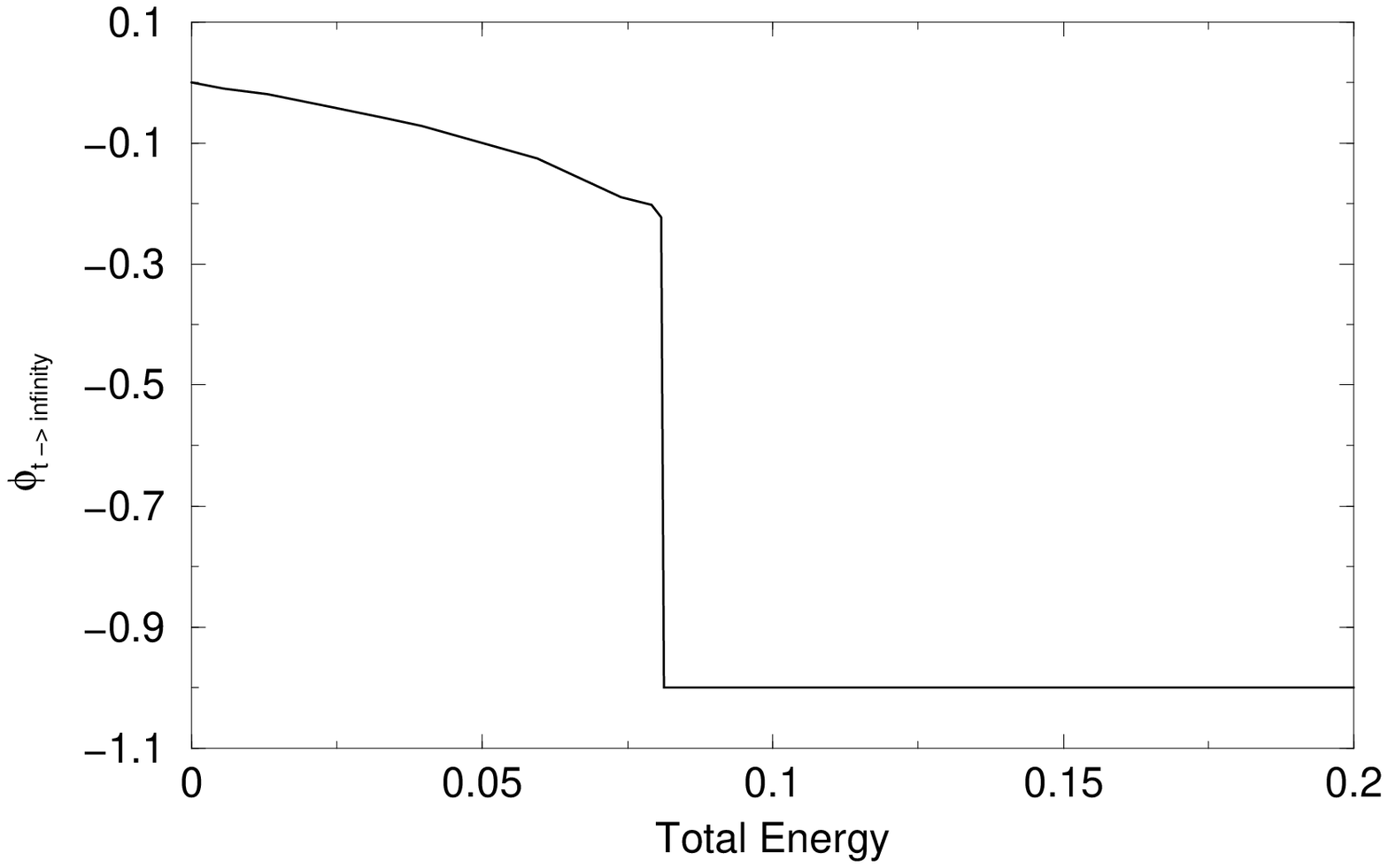}}
\end{center}}
\parbox[t]{7cm}
{{\small
FIG. 6: The scalar fluctuation $\langle A^2\rangle$ showing unstable
behavior, parameters as in Fig.~4.
}}
\hspace{0.5cm}\parbox[t]{7cm} {{\small FIG. 7: Asymptotic value of
$\phi$ as a function of the total energy with $m=\lambda=1.0$. Notice,
that $\phi_{\infty}=-m/\lambda=-1.0$ since the energy density is
larger than the threshold. } }}
\end{center}
\vspace{0.5cm}

Although the dynamics breaks down at one loop order, we can
nevertheless consider what will happen as the evolution proceeds into
the non-perturbative regime.  We are helped by the sum rule
(\ref{susysumrule}) for the masses of the fields which states that
$\calm_A^2 + \calm_B^2 = 2\calm_\psi^2$ (we mention once again that
this relation holds both for out of equilibrium unbroken and broken
supersymmetry).  First, we note that the one loop results satisfy this
relation, only with the caveat that the squared mass for $A$ is
negative.  Experience with non-perturbative
dynamics\cite{dynmaxwell,spindecomp,metric,spinodal} in purely scalar
field theories provides clues as to what to expect.  The main point is
that the resulting growth of fluctuations of $A$ due to the
instability will provide a contribution tending to increase the
effective mass of the field toward zero (and at the same time
decreasing the $B$ field mass toward zero).  The end result will be
massless fields $A$ and $B$.  This is, in fact, the only possible
stable solution of the sum rule given massless fermions.

We therefore expect that the concave portion of the classical
potential will become flattened out by non-perturbative effects due to
a spinodal instability in the field $A$.  This flattening has been
seen in studies of scalar $\lambda \Phi^4$ models in the large $N$
limit of an $O(N)$ field\cite{metric} and using the self-consistent
Hartree approximation for a single scalar field\cite{spinodal}.  These
studies verified that the flattening of the potential is related to
the dynamical approach to the Maxwell construction of the true free
energy of the system\cite{dynmaxwell,gruren}.

It is worthwhile to discuss the processes that lead to the transition
between the two dynamical regimes.

If the initial energy is sufficient for the system entering the
spinodal region, Eq.~(\ref{spinodalregion}), the region between a and
b displayed in Fig.~1, the instability of the low momentum modes of
the field ${\cal A}$ comes into play.  This spinodal energy density is
given by $\rho_s=m^4/18 \lambda^2$, marked by d in Fig.~1, and if we
start the motion with $\phi(0)> 0$, the corresponding initial
amplitude is $\phi_s=m(-1+\sqrt{5/3})/\lambda$. If $\phi(0)$ exceeds
$\phi_s$ only slightly the system enters the unstable region only for
a short time, and the instability does not build up. At late times the
classical field has transferred energy to the quantum modes and no
longer enters the spinodal region; it again ends up in the regime of
stationary oscillation.

At higher excitations the time spent by the system in the spinodal
region increases and after a few oscillations the system is trapped
there. The unstable low momentum modes of the field ${\cal A}$ grow
indefinitely and the classical amplitude tends to
$\phi_\infty=-m/\lambda$, i. e., at the maximum of the classical
potential.  In the phase diagram in Fig.~7 we have displayed the final
value of $\phi$ in dependence of the total energy. The first part of
the curve indicates that for low energies $\phi_\infty$ ends up near
the minimum of the classical potential as described above. At some
initial value $\phi(0)=\phi_{\rm crit}$ we find a rather sharp
transition into the unstable regime, $\phi_\infty=-m/\lambda$.

This dynamics continues for energies above the energy density of the
maximum of the potential $V_{\rm max}=m^4/8\lambda$, and to the
highest excitations we have considered. Above $\rho=V_{\rm max}$ the
oscillations of the system extend into the region of the second
supersymmetric minimum $\phi=-2m/\lambda$ and back again to $\phi=0$.
With each oscillation the system enters the spinodal region and the
instability can build up uninhibited. The result is that the system
again ends up at $\phi_\infty=-m/\lambda$.  This final state is not a
stationary state in the true sense, however; while the classical
amplitude approaches $\phi=-m/\lambda$ the fluctuations of the field
${\cal A}$ grow indefinitely, see Fig.~6.

\subsection{Fields near the local maximum of the potential}

We can analytically solve the field equations (\ref{oneloopeom}),
(\ref{Aexp}) and (\ref{modesc}) for the order parameter near the local
maximum of the potential $ \phi = -{m \over \lambda} $.

We set,
\be
\phi(t) = -{m \over \lambda} + \Delta(t)
\ee
and we will restrict to times where $ \Delta(t) \ll m $. 

Eqs.(\ref{modesc}) for $ f_A(t) $ thus becomes
$$
 {\ddot f}_A(t) + \left[k^2 -\frac{m^2}{2} + \frac32 \, \lambda^2 \,
 \Delta^2(t) \right]f_A(t)=0
$$
showing that the $A$-modes with $k< m/\sqrt2 $ are growing
exponentially as $ e^{t \sqrt{\frac{m^2}{2}-k^2}} $. Therefore, the
quantum fluctuations $ \langle A^2 \rangle $ [see eq.(\ref{afluct})] 
will be dominated by these spinodally unstable low-$k$ modes and  
will grow as
$$
\langle A^2\rangle(t) = \intk \frac{|f_A(t)|^2}{2\omega_{ A 0}}
\buildrel{mt \gtrsim 1 }\over = { e^{\sqrt2 \; m \, t} \over
(m\,t)^{3/2} } D(t) \; ,
$$
where the function $ D(t) $ is the order one for late times.
The equation of motion for the zero mode (\ref{oneloopeom}) takes the
form
\be \label{cercapun}
{\ddot \Delta}(t) + \left\{ -\frac{m^2}{2} + \frac12 \, \lambda^2
\left[\Delta^2(t) + 3 \langle A^2 \rangle(t) + \langle B^2 \rangle(t)
\right] \right\} \Delta(t) + \frac12 \, \lambda \langle \overline{\psi} \psi
\rangle(t) =0\; .
\ee
Since as $t$ grows the zero mode approaches  $ -{m \over \lambda} $, $
\Delta(t) \to 0 $ and $  \langle A^2 \rangle(t) \to \infty
$. Therefore, in order eq. (\ref{cercapun}) keep valid, the product
$$
\langle A^2 \rangle(t) \; \Delta(t) = G(t) \; ,
$$
must stay bounded. Moreover, it must fulfill
\be \label{consi}
G(t) = - {1 \over 3 \, \lambda } \langle \overline{\psi} \psi \rangle(t)
\; ,
\ee
a relation which is satisfied by the dynamics in the numerical 
simulations, see Fig.~8.

\noindent
\parbox{14.8cm}{
%\parbox[t]{8cm}
{\begin{center}
\mbox{\epsfxsize=7.5cm\epsfbox{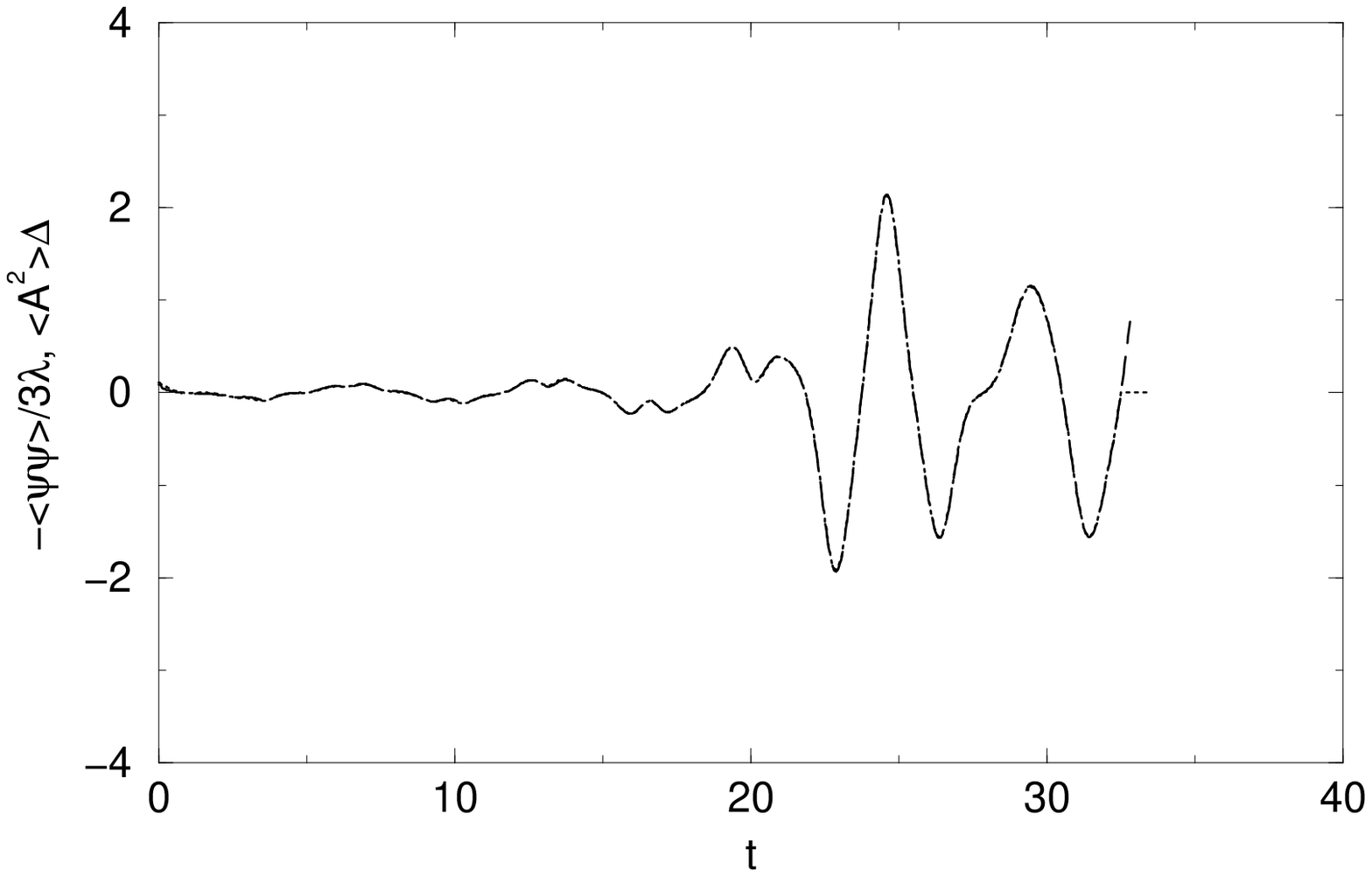}}
\end{center}}
\hspace{.0cm} {{\small FIG. 8: Numerical check of $\langle A^2
\rangle\Delta=-\langle\bar\psi\psi\rangle/3\lambda$, dashed line:
$\langle A^2\rangle\Delta$, dotted line:
$-\langle\bar\psi\psi\rangle/3\lambda$.  }}}

\vspace{0.5cm}

\section{Large $N$ extension of the Wess-Zumino model} \label{sec3}
\label{nsusy}

Having seen that the self-consistent one-loop approximation to the
Wess-Zumino model breaks down due to the spinodal instability, we now
turn to an extension of the model for which we can follow the
evolution into the non-perturbative domain.  We couple the Wess-Zumino
superfield to $N$ chiral superfields satisfying an internal $O(N)$
symmetry.  By taking the large $N$ limit, we arrive at a
semi-classical supersymmetric model for which we can numerically solve
for the exact and fully non-perturbative field dynamics.

\subsection{Model and equations of motion}

Our $O(N)$ extension of the Wess-Zumino model 
consists of a chiral superfield multiplet $S_0=(A_0,B_0;\psi_0;F_0,G_0)$, 
which acts as a singlet under $O(N)$, coupled to $N$ chiral superfields 
$S_i=(A_i,B_i;\psi_i;F_i,G_i)$ with $i = 1\dots N$ and which transform 
as a vector under $O(N)$.  The superpotential has the form
\begin{eqnarray}
{\cal L} &=&  \frac12 M S_0 \cdot S_0 + \frac{\kappa}{6\sqrt{N}} 
S_0 \cdot S_0 \cdot S_0
+ \frac12 m \sum_{i=1}^N  S_i \cdot S_i +  \frac{\lambda}{2\sqrt{N}}  
\sum_{i=1}^N S_0 \cdot S_i \cdot S_i \; .
\label{susyNlagrangian}
\end{eqnarray}
Notice that the model considered in Ref.\cite{zanon} is a special case
of ours for $ m = \kappa = 0 $. 
Again, we expand in terms of the component fields and eliminate
the auxiliary fields via their equations of motion.  To allow 
for a consistent large $N$ limit, the expectation value of $A_0$ 
is taken to be of order $\sqrt{N}$.  We set
\begin{equation}
\langle A_0 \rangle = \sqrt{N}\phi \, \, , \; \langle B_0 \rangle = 0
\; .
\end{equation}
The latter condition, which amounts to a choice of initial conditions,
is chosen because it significantly simplifies the equations of motion.
We assume that the initial state satisfies the $O(N)$ symmetry 
which requires that $\langle A_i \rangle = \langle B_i \rangle = 0$.
Given the $O(N)$ symmetry, it is convenient to define fields $A$,
$B$, and $\psi$ such that $\sum_i A_i A_i = N A^2$, 
$\sum_i B_i B_i = N B^2$, and 
$\sum_i \overline{\psi}_i\psi_i = N \overline{\psi}\psi$. 
In taking the large $N$ limit, particular care must be given
to terms of the form $\sum_i A_i B_i$ which turn out to be of order
$\sqrt{N}$. This is most easily seen by computing the squared quantity
$\sum_i \sum_j A_i B_i A_j B_j$; the resulting delta function $\delta_i^j$
contributes a factor $1/N$ which cancels one of the factors of $N$ coming
from the summations.  The Lagrangian to leading order in $N$ is
\begin{eqnarray}
\frac{{\cal L}}{N} &=& \frac12 \partial_\mu \phi \partial^\mu \phi 
+\frac12 \partial_\mu A \partial^\mu A
+\frac12 \partial_\mu B \partial^\mu B 
-\frac12 \phi^2 \left(M + \frac12 \kappa \phi \right)^2
\nonumber\\
&&-\frac12 \left(m + \lambda \phi\right)^2 \left(A^2 + B^2\right)
-\frac{\lambda}{2}\left[M\phi + \frac12 \kappa \phi^2 
+ \frac14 \lambda \left(A^2 - B^2\right)\right] \left(A^2 - B^2\right)
\nonumber\\
&&+\frac{i}{2}\bar\psi\Slash\partial\psi-\frac 1 2 m\bar\psi\psi
-\frac{1}{2}\lambda \phi \bar\psi\psi \; .
\label{largeNlagrangian}
\end{eqnarray}

Note that the fluctuations of the supersymmetry multiplet $S_0$ do not
appear to this order in $N$, with only the mean field $\phi$
contributing to the Lagrangian.  The consequence is that the
Lagrangian of this theory is equivalent to a supersymmetric $O(N)$
field $S_i$ coupled to a fully classical background zero mode $\phi$.
This is a consequence of the semi-classical nature of the large $N$
approximation.

The equation of motion for the mean field $\phi$ is found to be
\begin{eqnarray}
\label{mean}
\ddot{\phi} &+& (M+\kappa \phi) \left[M\phi + \frac12 \kappa \phi^2
+ \frac12 \lambda \langle A^2-B^2 \rangle \right] \nonumber \\
&+& \lambda(m+\lambda\phi) \langle A^2+B^2 \rangle 
+ \frac12 \lambda \langle \overline{\psi}\psi \rangle = 0
\; ,
\end{eqnarray}
while the fermion obeys
\begin{equation}
\left[i \Slash\partial - (m+\lambda\phi)\right] \psi = 0 \; ,
\end{equation}
and the mode functions from which the scalar fluctuations are 
built satisfy
\begin{eqnarray}
\ddot{f}_A + k^2 f_A + (m+\lambda\phi)^2 f_A 
+ \lambda\left[M\phi+\frac12 \kappa \phi^2 + \frac12 \lambda \langle A^2 
- B^2\rangle \right] f_A &=& 0 \; , \\
\ddot{f}_B + k^2 f_B + (m+\lambda\phi)^2 f_B 
- \lambda\left[M\phi+\frac12 \kappa \phi^2+ \frac12 \lambda \langle A^2 
- B^2\rangle \right] f_B &=& 0 \; .
\end{eqnarray}
The integrals for the fluctuations are constructed just as in the
ordinary Wess-Zumino model via Eqs.~(\ref{afluct})--(\ref{psifluct}).
It is convenient to define the masses
\begin{eqnarray}
\label{defmmin}
\calm_\psi & \equiv & m+\lambda\phi \; , \\
\calm_-^2 & \equiv & \lambda\left[M\phi+\frac12 \kappa \phi^2
+ \frac12 \lambda \langle A^2 - B^2\rangle \right] \; ,
\label{m-eqn}
\end{eqnarray}
such that 
\begin{eqnarray}
\label{lnflmass1}
\calm_A^2 &=& \calm_\psi^2 + \calm_-^2 \; , \\
\label{lnflmass2}
\calm_B^2 &=& \calm_\psi^2 - \calm_-^2 \; .
\end{eqnarray}
We see again that the supersymmetry sum rule 
$\calm_A^2 + \calm_B^2 -2\calm_\psi^2 = 0$ 
is automatically satisfied.

The equation for $\calm_-^2$, (\ref{m-eqn}), plays the role of a gap
equation expected in the large $N$ framework, since the fluctuations
appearing in the integrals on the right hand side depend upon the
masses $\calm_A$ and $\calm_B$ which in turn depend upon $\calm_-$.
This becomes more explicit in the fully renormalized form found in
Appendix B.  We also note that $\calm_-^2$ can be either positive or
negative, with the consequence that either or both $\calm_A^2$ and
$\calm_B^2$ can become negative during the evolution.
 
The energy density is 
\bea \nonumber
\cale&=&\frac12 \dot\phi^2 + \frac12 \left(M\phi
+\frac{\kappa}{2}\phi^2\right)^2
\\  &&
+\frac12 \langle \dot A^2+k^2A^2+\calm_A^2(t) A^2\rangle
+\frac12 \langle \dot B^2+k^2B^2+\calm_B^2(t) B^2\rangle
\\ &&\nonumber 
-\frac{\lambda^2}{8}\left(\langle A^2-B^2\rangle\right)^2
+\frac{1}{2}\langle\bar\psi\left(-i\vec\gamma \cdot \vec\nabla
+\calm_\psi(t)\right)\psi\rangle
\pkt
\eea

Again the equations of motion and the energy density have to be
renormalized. Despite the non-perturbative nature of the large $N$
limit, one of its important properties is that it is possible to
consistently renormalize the theory. The details are presented in
Appendix B. We separate out the divergent terms so that they may be
treated analytically, leaving the finite parts to be included in
numerical simulations.  This allows us the freedom to choose a
regularization scheme without regard to constraints of the numerical
simulations.  We then use dimensional regularization to define the
counterterms.  Again the structure of the finite equations remains
essentially the same as the one of the bare equations, with divergent
expectation values replaced by finite ones, and with finite
corrections to the masses and couplings.  We find consistency with the
renormalization of the large-$N$ equilibrium theory.

\subsection{The supersymmetry in the large $N$ limit}

Examining the leading order Lagrangian, Eq.~(\ref{largeNlagrangian}),
the question arises as to whether taking the large $N$ limit is
consistent with supersymmetry.  One might be particularly concerned
since the singlet superfield $S_0$ appearing in
(\ref{susyNlagrangian}) is represented only by the single scalar zero
mode $\phi = \langle A_0 \rangle/\sqrt{N}$ and one might expect that
it would be necessary to have a corresponding fermion field into which
$\phi$ may transform.  To clarify this point we examine the
transformation properties of $A_0$.  Under supersymmetry, $A_0$
transforms as
\begin{equation}
A_0^{'} = A_0 + \delta A_0 = A_0 + \overline{\zeta} \psi_0 \; , 
\label{a0trans}
\end{equation}
where $\zeta$ is the Grassman supersymmetry transformation parameter.  
Taking the expectation value of Eq.~(\ref{a0trans}) yields the
transformation law for $\phi$:
\begin{equation}
\phi^{'} \equiv \langle A_0^{'} \rangle/\sqrt{N} 
= \langle A_0 \rangle/\sqrt{N} 
+ \langle \overline{\zeta} \psi_0 \rangle/\sqrt{N} 
= \phi + \overline{\zeta} \langle \psi_0 \rangle/\sqrt{N} 
= \phi \; ,
\label{phitrans}
\end{equation}
where the last equality follows from the vanishing expectation value
of $\psi_0$.  We therefore see that $\phi$ is invariant under
supersymmetry transformations.  This is consistent with the treatment
of $\phi$ as a classical background zero mode coupled to the
supersymmetric $O(N)$ multiplet.

To leading order, it is no longer necessary to consider the
transformation properties of the $O(N)$ singlet fields corresponding
to the superfield $S_0$ as none of these fields appear in our large
$N$ Lagrangian (\ref{largeNlagrangian}).  The remaining
transformations are:
\begin{eqnarray}
\delta \phi &=& 0 \; , \\
\delta A &=& \overline{\zeta} \psi \; , \\
\delta B &=& \overline{\zeta} \gamma_5 \psi \; , \\
\delta \psi &=& \left(i\Slash\partial + m + \lambda \phi\right)
\left(A+\gamma_5 B\right)\zeta \; , \\
\delta \overline{\psi} &=& \overline{\zeta}
\left[-i\Slash\partial \left(A-\gamma_5 B\right) 
+ \left(m+\lambda \phi\right) \left(A+\gamma_5 B\right)\right] \; . 
\end{eqnarray}
Through use of the equations of motion for $A$, $B$, and $\psi$, it is
straightforward to show that the variation of the Lagrangian
(\ref{largeNlagrangian}) vanishes under these supersymmetry
transformations up to a total derivative.  Therefore, we see that to
this order, the Lagrangian is completely supersymmetric with $\phi$
acting as a classical background field.

\subsection{Results and interpretation}

As in the ordinary Wess-Zumino model, we find two distinct phases, one
at low energy densities and one at high energy densities.

Unlike the ordinary Wess-Zumino model, however, this model has
perturbative vacua for which the fermion masses vanish. 
  
In the high energy density phase, the time evolution leads to a final
state formed by a cloud of massless particles (both bosons and
fermions) around a minimum $ \phi = -m/\lambda $ of the effective
potential, degenerate with the tree level minimum.  This
nonperturbative minimum defines a new sector of the theory.  [The
perturbative sector is formed by particles around the perturbative
ground state $ \phi = 0 $.]  Since this nonperturbative minimum at $
\phi = -m/\lambda $ is degenerate with the perturbative minimum, it is
invariant under supersymmetry.  Hence, supersymmetry is not broken at
this nonperturbative minimum. In addition, the fact that all masses
vanish at the new minimum $ \phi = -m/\lambda $ supports the
supersymmetric character of this point.  In addition, we choose $O(N)$
invariant states since $\langle A_i \rangle = \langle B_i \rangle = 0$
for all times.

However, the final state obtained by real time evolution for large
energy density is formed by a cloud of massless particles (both bosons
and fermions) around this new nonperturbative minimum $ \phi =
-m/\lambda $. This state indeed breaks supersymmetry since the energy
is distributed differently among the fermions and bosons on the top of
this zero-energy nonperturbative ground state due to their differing
statistics.  This is analogous to which is known from equilibrium
studies at finite temperature where a thermal gas of particles is
around a perturbative supersymmetric vacuum.

In summary, for large energy density the system goes to a
nonperturbative minimum at $ \phi = -m/\lambda $ which acts as an
attractor to the dynamics and where a kind of Goldstone phase develops
with all particles, bosons and fermions, being massless.
Interestingly, these phenomena happens for the highest energy
densities tested.  The reason that keeps all particles massless is
that supersymmetry results in cancellations in the contributions to
the effective masses of the fields, see eqs.(\ref{m-eqn}) --
(\ref{lnflmass2}).

The situation here looks similar to the case where there is no
symmetry restoration at high energies and massless Goldstone bosons
are therefore present \cite{sw,varios}.  While individual field
fluctuations may become large as the energy is increased, the net
contribution of the fluctuations does not grow. Thus, arbitrarily high
energy densities need not yield the mass corrections which ordinarily
lead to symmetry restoration in spontaneously broken $O(N)$-theories
\cite{dynmaxwell}.

We now examine each phase in detail.

The low energy, massive phase occurs in much the same way as in the
ordinary Wess-Zumino model and is depicted in Fig.~9.  with the
corresponding masses shown in Fig.~10.  However, because of the more
complicated structure of the potential in the large $N$ model, the
range of energy densities for which the low energy density phase
persists depend on the initial conditions.  One can distinguish two
cases:
\begin{enumerate}
\item $\phi_0 > -M/\kappa$ and $m/\lambda < M/\kappa$, or $\phi_0 <
  -M/\kappa$ and $m/\lambda > M/\kappa$.  In either of these two
  cases, the zero mode $\phi$ begins on the same side of its potential
  as the new nonperturbative supersymmetric minimum.  This results in
  a relatively low critical energy density for reaching the high
  energy density phase.
\item $\phi_0 > -M/\kappa$ and $m/\lambda > M/\kappa$, or $\phi_0 <
  -M/\kappa$ and $m/\lambda < M/\kappa$.  In these cases, there is an
  additional potential barrier between the initial value of $\phi$ and
  the new nonperturbative supersymmetric minimum.  As a result, the
  transition to the high energy density phase occurs at a higher
  critical value of the energy density.
\end{enumerate}

\noindent
\begin{center}
\parbox{14.8cm}{
\parbox[t]{7cm}
{\begin{center}
\mbox{\epsfxsize=6.5cm\epsfbox{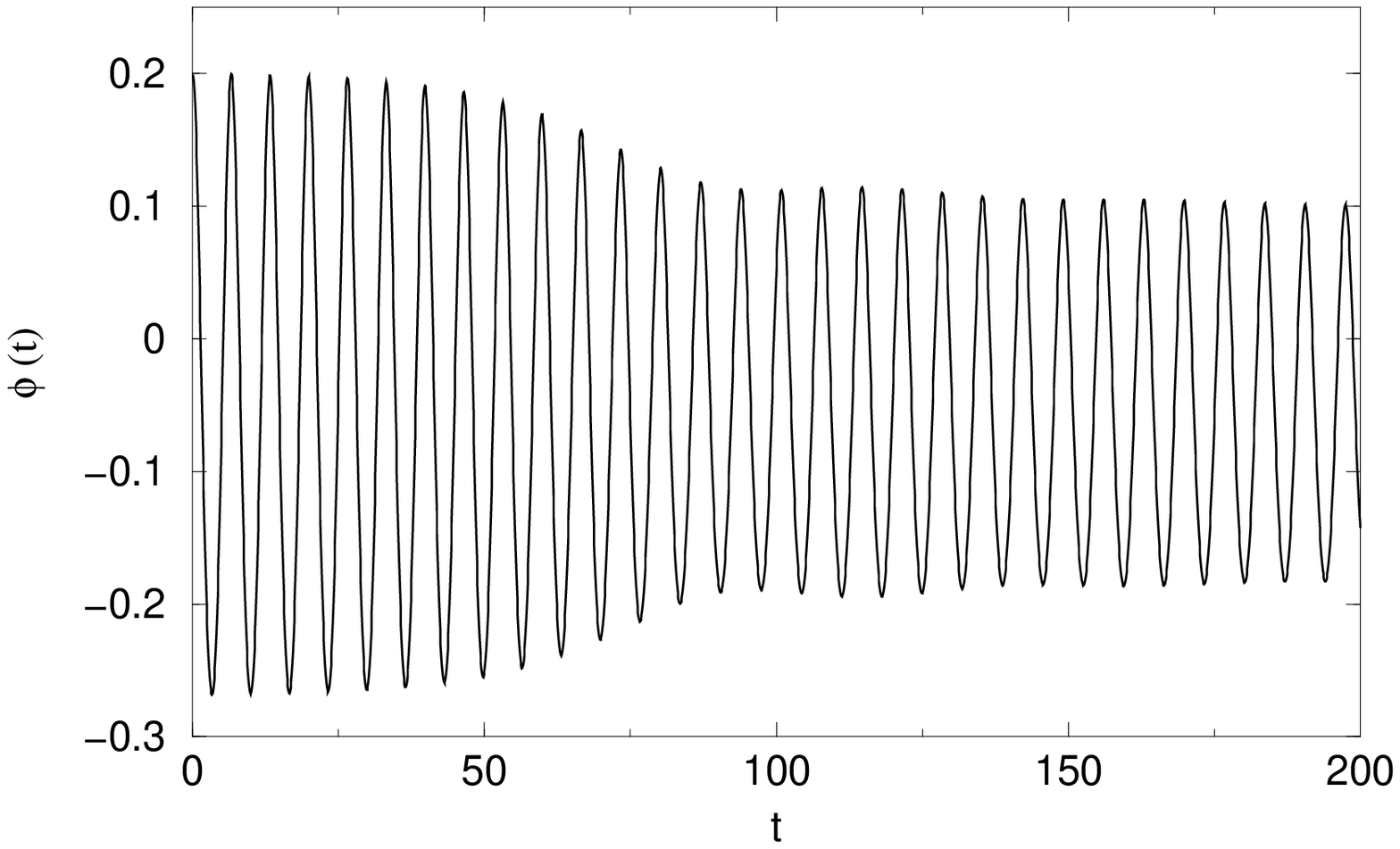}}
\end{center}}
\hspace{.0cm}\parbox[t]{7cm}
{\begin{center}
\mbox{\epsfxsize=6.5cm\epsfbox{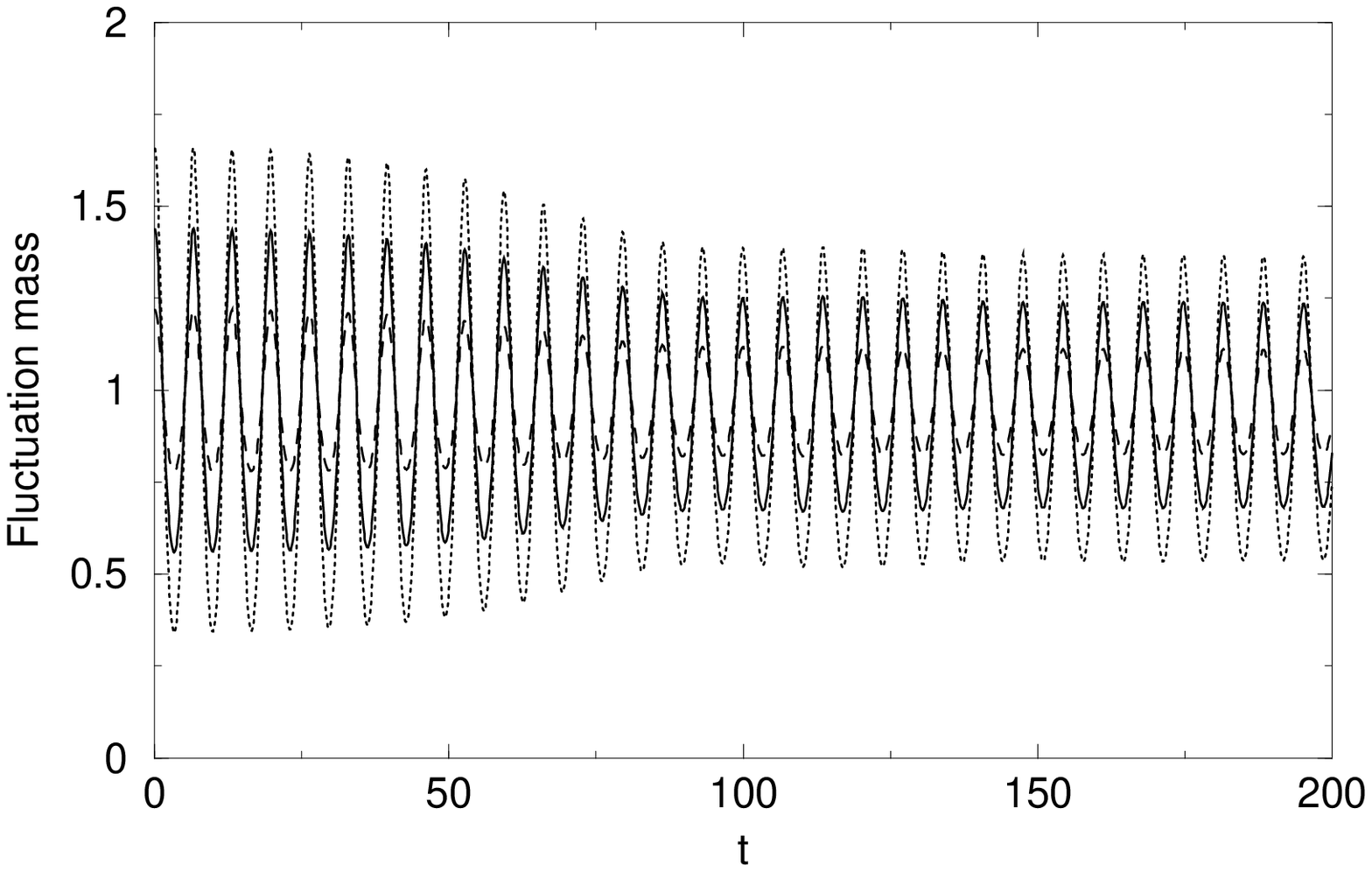}}
\end{center}}
\parbox[t]{7cm}
{{\small
FIG. 9: Zero mode evolution showing the
low energy density phase for $m=M=1.0$, $\kappa=\lambda=1.0$ and $\phi(0)=0.2$.
}}
\hspace{0.5cm}\parbox[t]{7cm}
{{\small FIG. 10: The effective squared masses; 
solid line: $\calm_\psi^2(t)$, dotted line: $\calm_A^2(t)$, 
dashed line: $\calm_B^2(t)$; the parameters 
as in Fig.~9.
}
}}
\end{center}

\vspace{0.5cm}

\noindent
\begin{center}
\parbox{14.8cm}{
\parbox[t]{7cm}
{\begin{center}
\mbox{\epsfxsize=6.5cm\epsfbox{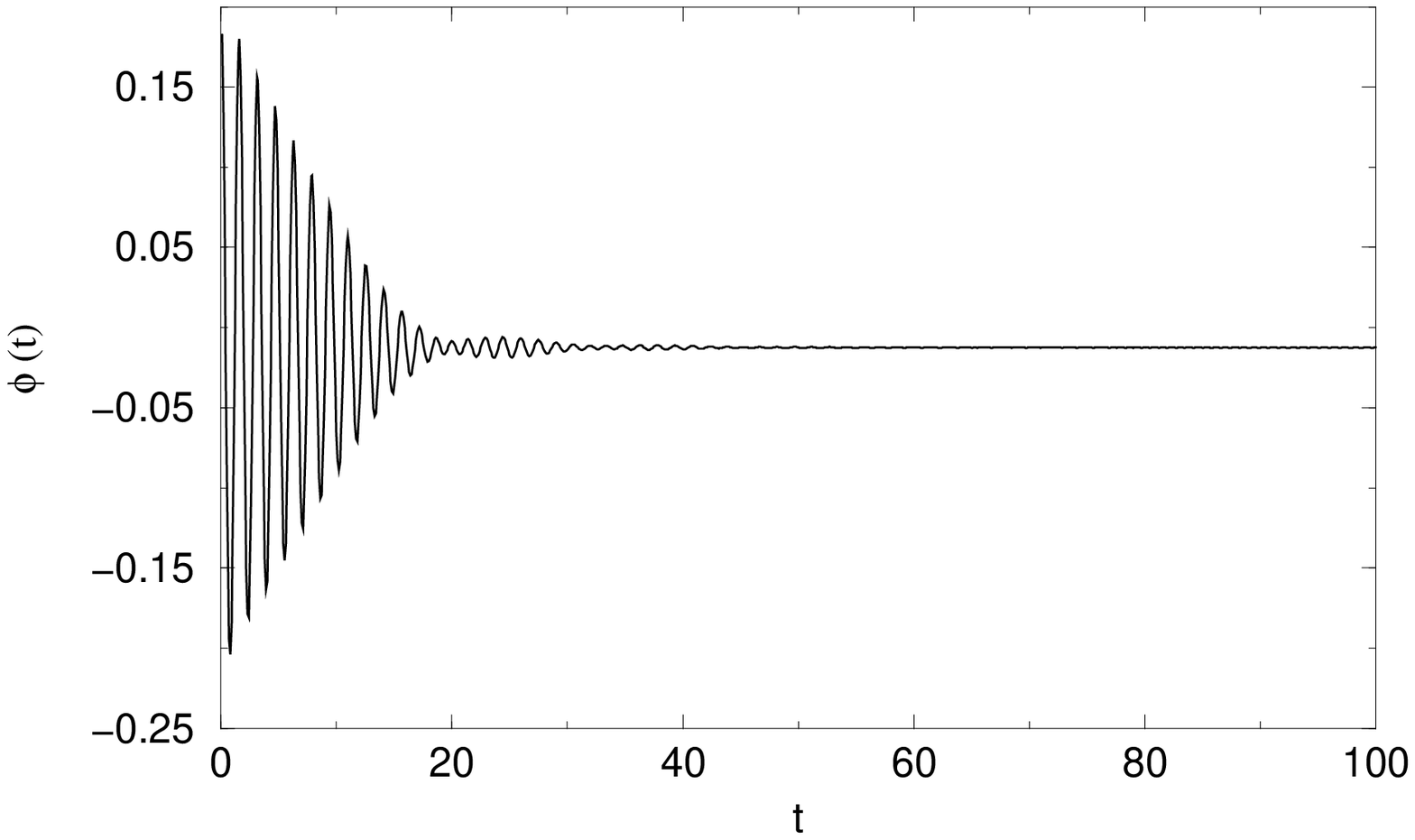}}
\end{center}}
\hspace{.0cm}\parbox[t]{7cm}
{\begin{center}
\mbox{\epsfxsize=6.5cm\epsfbox{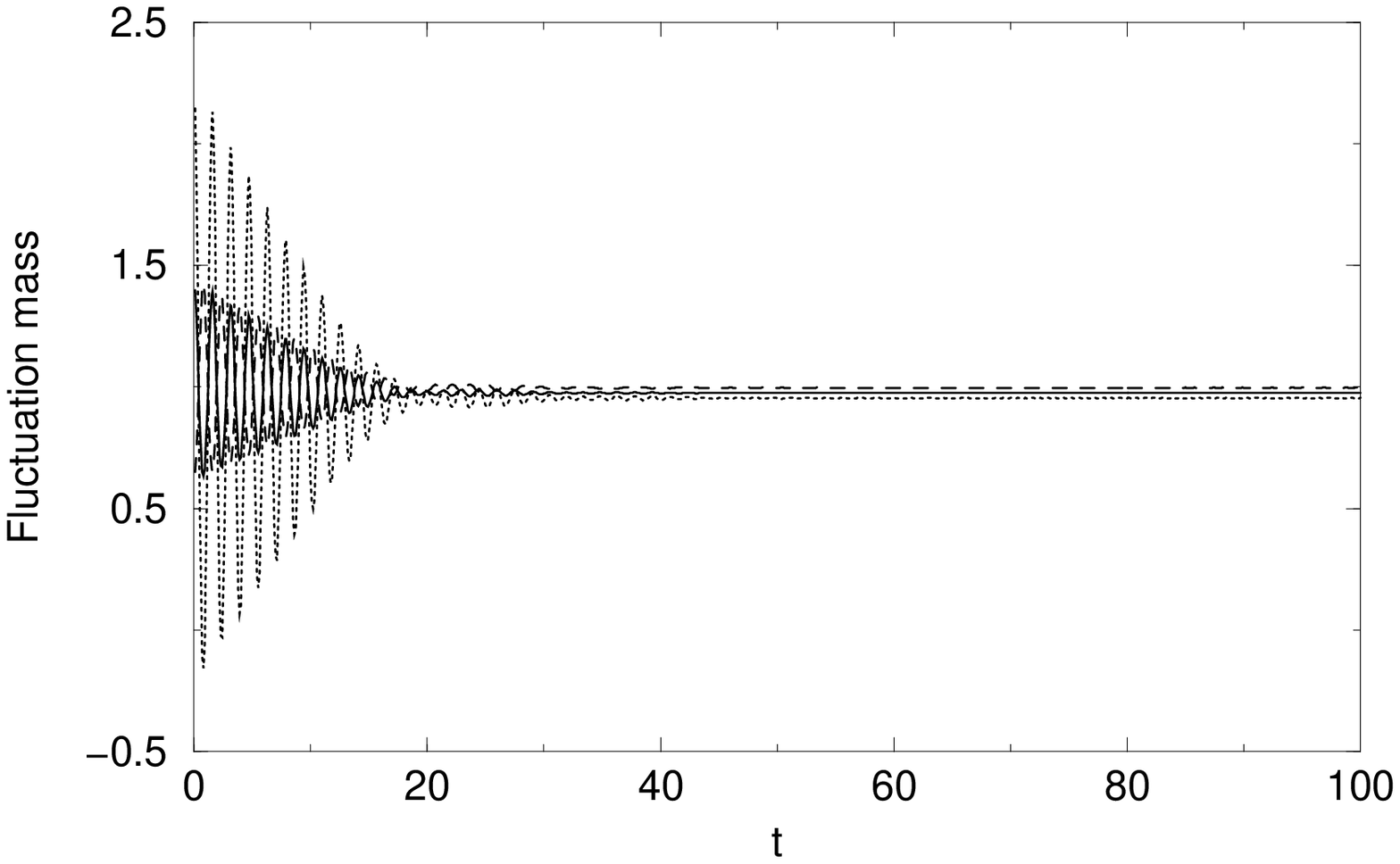}}
\end{center}}
\parbox[t]{7cm}
{{\small
FIG. 11: Zero mode evolution showing the
low energy density phase for $m=1.0$, $M=4.0$, $\kappa=\lambda=1.0$ and $\phi(0)=0.2$.
}}
\hspace{0.5cm}\parbox[t]{7cm}
{{\small FIG. 12: The effective squared masses; 
solid line: $\calm_\psi^2(t)$, dotted line: $\calm_A^2(t)$, 
dashed line: $\calm_B^2(t)$; parameters 
as in Fig.~11.
}
}}
\end{center}

\vspace{0.5cm}

A second example of low energy density evolution is shown in Figs.~11
and 12 for which the masses are such that $\phi$ may efficiently decay
into $A$ and $B$ particles, leading to apparent dissipation.  We see
that $\phi$ decays and settles at a point near the classical minimum
at $\phi = 0$.  The shift of the finite density minimum from this
vacuum value is due to the growth of the fluctuations of the fields
$A$, $B$, and $\psi$, which are also responsible for the deviation of
the masses from the supersymmetry value $m$.

An example of the high energy density phase is depicted in Fig.~13
with the masses in Fig.~14.  The evolution begins with large
oscillations of $\phi$ over the entire classically allowed range of
evolution.  During this initial period, the field fluctuations
$\langle A^2 \rangle$ and $\langle B^2 \rangle$ grow.  After a
relatively short period of time, the mean field settles down precisely
to the point $\phi = -m/\lambda$.  The result is that the $N$ fermions
become massless.  As in the one-loop Wess Zumino case, this is an
attractor state at high energy densities.  However, in the present
model the massless state is stable.  A look to the effective potential
\cite{BCdVH} shows that $ \phi = -m/\lambda $ is a zero-energy minimum
of the effective potential. This minimum is degenerate with the
perturbative vacuum $ \phi = 0 $.

\noindent
\begin{center}
\parbox{14.8cm}{
\parbox[t]{7cm}
{\begin{center}
\mbox{\epsfxsize=6.5cm\epsfbox{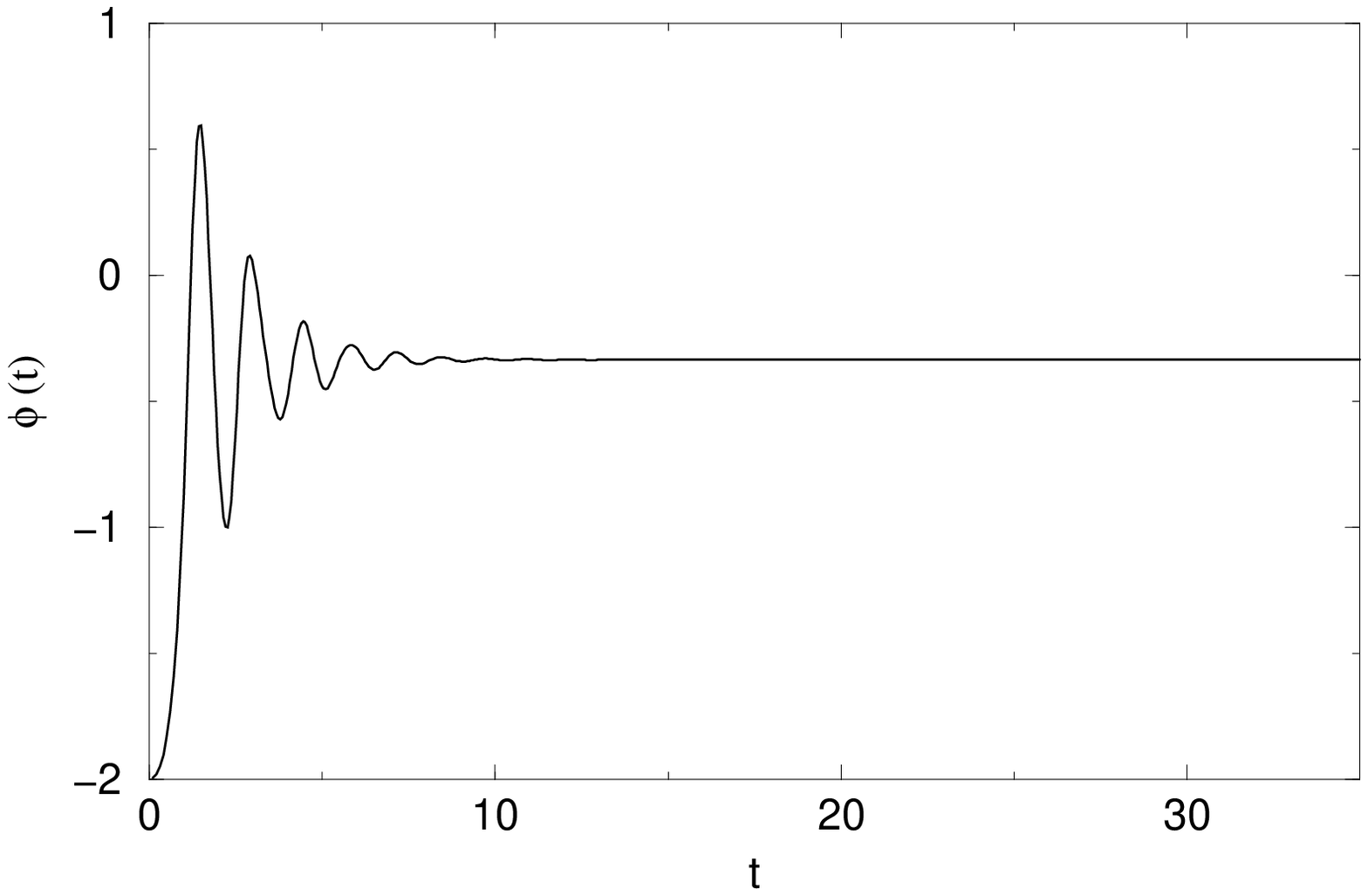}}
\end{center}}
\hspace{.0cm}\parbox[t]{7cm}
{\begin{center}
\mbox{\epsfxsize=6.5cm\epsfbox{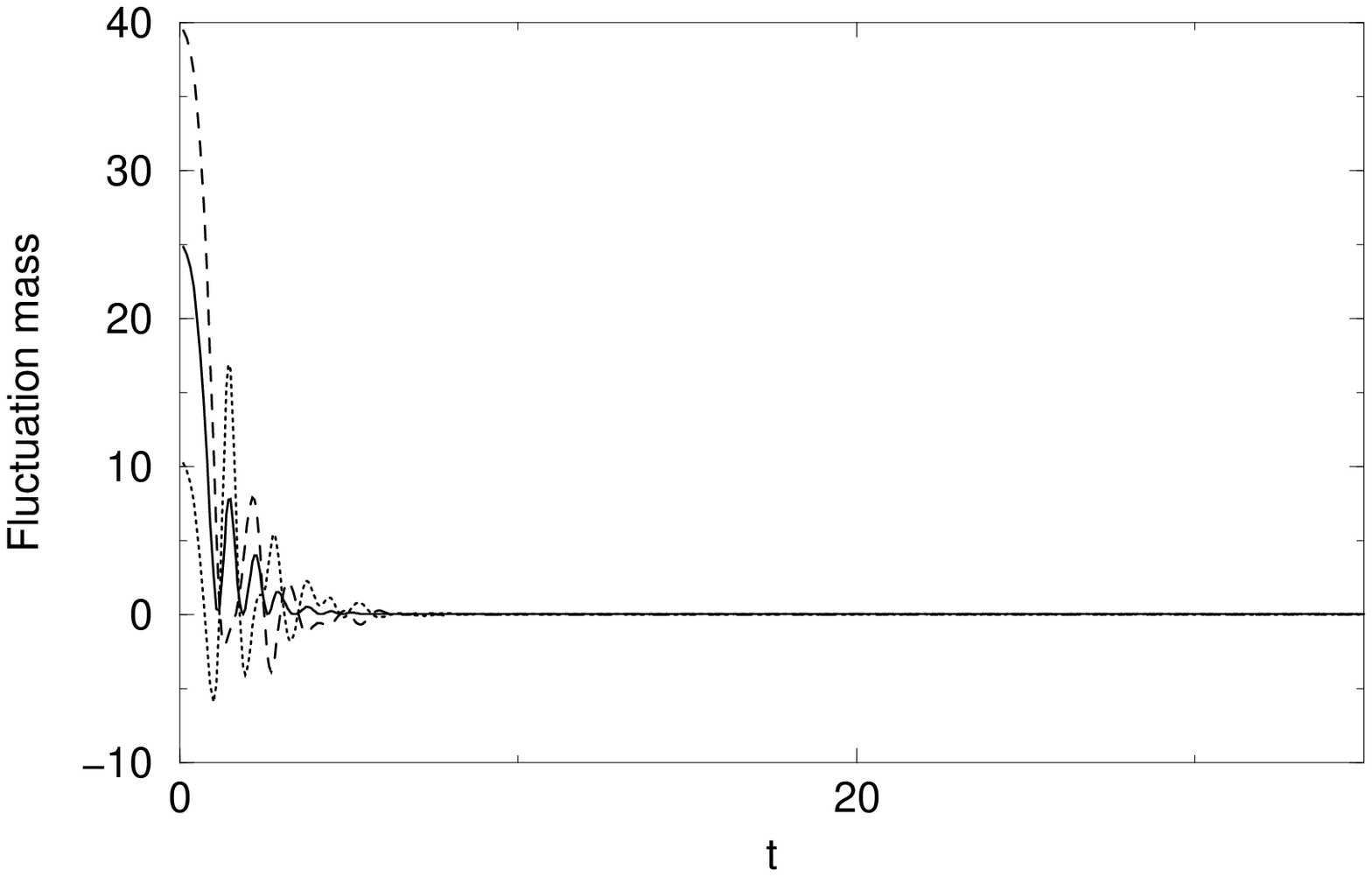}}
\end{center}}
\parbox[t]{7cm}
{{\small
FIG. 13: Zero mode evolution showing the
high energy density phase for $m=1.0$, $M=4.0$, $\kappa=2.0$, $\lambda=3.0$ 
and $\phi(0)=-1.5$.
}}
\hspace{0.5cm}\parbox[t]{7cm}
{{\small FIG. 14: The effective squared masses; 
solid line: $\calm_\psi^2(t)$, dotted line: $\calm_A^2(t)$, 
dashed line: $\calm_B^2(t)$; parameters 
as in Fig.~13.
}
}}
\end{center}

\vspace{0.5cm}

We therefore find that the system reaches a non-perturbative state for
which each of the $O(N)$ fermion and boson fields is massless and for
which the vacuum energy identically vanishes as well indicating the
presence of a new non-perturbative supersymmetric minimum at $\phi =
-m/\lambda$.  The fact that the massless state is an attractor is of
great importance if one considers the model from a cosmological
perspective.  What it indicates is that the initial evolution of the
system in the early universe can determine the ultimate vacuum state
of the system, providing an effective means of selection between vacua
which are otherwise degenerate in energy.

\section{Adding soft breaking terms} \label{sec4}

One question that might be raised is what are the consequences of
adding additional mass terms to the model which softly break the
supersymmetry.  This is particularly interesting because, as we have
seen, the dynamics leads to massless fermions and, in the case of the
$O(N)$ model, massless bosons as well.  Through soft supersymmetry
breaking, it may be possible to introduce small masses to these final
state particles, which would be determined not by the scale of
supersymmetry, rather by some lower scale (e.g.., the electroweak
scale) at which the soft supersymmetry breaking terms arise.

To provide an example case, we introduce soft supersymmetry breaking
to the $O(N)$ model via a scalar mass $m_s$ for the $A$ and $B$ fields
such that Eqs.~(\ref{lnflmass1}) and (\ref{lnflmass2}) become
$\calm_A^2 = \calm_\psi^2 + \calm_-^2 + m_s^2$ and $\calm_B^2 =
\calm_\psi^2 - \calm_-^2 + m_s^2$.  Such terms break the supersymmetry
explicitly, while not producing any dangerous (i.e.  non-logarithmic)
divergences.  We plot the results for the case $m_s/m = 0.1$ in
Fig.~15.

\noindent
\parbox{14.8cm}{
%\parbox[t]{8cm}
{\begin{center}
\mbox{\epsfxsize=7.5cm\epsfbox{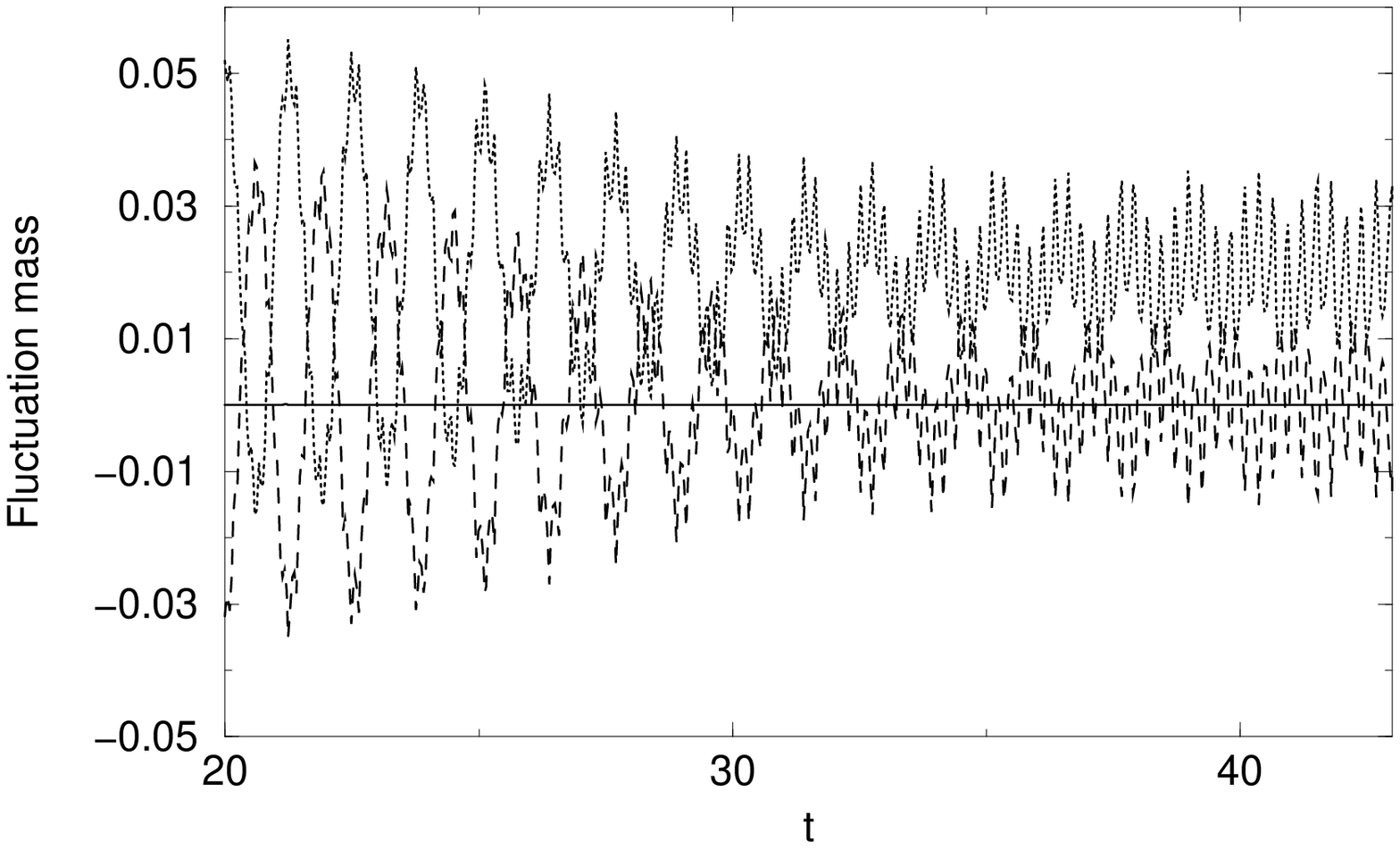}}
\end{center}}
\hspace{.0cm} {{\small FIG. 15: The effective field masses
in the high energy density phase including soft breaking masses for
the fields $A$ and $B$; solid line: $\calm_\psi^2(t)$, dotted line:
$\calm_A^2(t)$, dashed line: $\calm_B^2(t)$; the parameters are
$m=1.0$, $M=4.0$, $m_s=0.1$, $\kappa=2.0$, $\lambda=3.0$,
$\phi(0)=0.9$.  }}}

\vspace{.5cm}

The first thing to note is that we retain the high energy density
attractor solution.  As the new terms do not explicitly break the
$O(N)$ symmetry, there is necessarily one set of asymptotically
massless Goldstone bosons, in this case represented by the $B$ field.
However, these fields' superpartners are no longer massless as they
pick up contributions proportional to the soft supersymmetry breaking
mass scale $m_s$.  For $A$, this mass is given approximately by ${\cal
  M}_A \simeq \sqrt{2} \, m_s$, corresponding to ${\cal M}_{-} \simeq
m_s$ and ${\cal M}_{\psi} \simeq 0$.  For the fermions, there is a
see-saw mechanism producing the mass such that the leading order
contribution is proportional to $m_s^2/m$.  This produces fermion
masses which are suppressed by a factor of $m_s/m$ relative to their
massive bosonic superpartners.

\section{Discussion and conclusions} \label{sec5}
Let us first summarize our main results.
\begin{enumerate}
\item At low energy density, both the one loop and the large $N$
  Wess-Zumino models are found to allow for finite density
  non-equilibrium states based on supersymmetric vacua.  Finite
  density corrections break the mass degeneracy of the scalars and
  fermions somewhat, but do not lead to obvious massless fermions.  In
  the large $N$ model, such states are stable, albeit time dependent.
  They are also stable in the ordinary Wess-Zumino model to one loop
  order.
\item At high energy density, the dynamics of the Wess-Zumino model
  and its large $N$ extension leads to massless fermions.  These
  states are stable in the large $N$ model, but are highly unstable at
  one loop order in the ordinary Wess-Zumino model.
\item In the large $N$ model, the high energy density phase is
  characterized for arbitrarily high energy densities by all of the
  $O(N)$ vector fields, scalars and fermions, becoming massless, and
  by the vanishing of the total vacuum energy.  The phase is formed by
  a cloud of massless particles (both bosons and fermions) around a
  new nonperturbative supersymmetric minimum at $ \phi = -m/\lambda $.
\item The introduction of explicit soft supersymmetry breaking terms
  to the large $N$ Lagrangian results in masses for some of the $O(N)$
  vector fields.  In particular, we find that a soft supersymmetry
  mass of $m_s$ for the scalars induces a mass for the $A$ field
  $\calm_A = \sqrt2 m_s$ and a fermion mass of order $m_s^2/m$, while
  $B$ remains massless.
\end{enumerate}

We can visualize a number of consequences for cosmology and particle
physics.
\begin{enumerate}
\item It appears that we should expect massless fermions to occur in
  supersymmetric models of the early universe independent of whether
  the situation is one of equilibrium or far from equilibrium, or at
  least, if the universe is far from equilibrium, we should expect it
  to approach such a state.
\item If the vacuum energy is indeed zero as in the large $N$ case,
  then we have a convenient mechanism by which otherwise massive
  fermions (and possibly other fields as well) become massless.  This
  could have the effect of priming the system so that soft
  supersymmetry breaking terms can yield fields with masses much
  smaller than the overall scale of supersymmetry.
\item A final consequence of potential interest is that supersymmetry
  may protect continuous symmetries from being restored in the very
  early universe, as in our $O(N)$ model.  Such a result might
  alleviate the monopole problem, as without a symmetry restored
  phase, there would be no phase transition to produce such objects.
\end{enumerate}

We have only begun to examine these possibilities, but many of these
ideas can be pursued further using current techniques in out of
equilibrium quantum field theory.  One avenue of approach is to
examine the ordinary Wess-Zumino beyond one loop order by means of
mean field theory.  This may clarify what happens to the masses of the
$A$ and $B$ fields as well as having the potential to determine the
vacuum energy of the resulting non-perturbative state.  Also of
interest is the behavior of these models in an expanding universe, as
it would be important to see if a phase transition from the high
energy phase to the low energy density phase occurs when energy is
drained from the system via expansion.  The relation between the
expansion time scale and the relaxation time scale in the high energy
density phase may also play a significant role.

Of course, it would be beneficial to examine the detailed behavior of
gauge multiplets in order to begin to get a better understanding of
more realistic models of particle physics.  It would also be very
interesting to examine the possibility that neutrino masses could be
realistically generated via an analogous mechanism to that of fermion
mass generation in the softly broken supersymmetry model studied here.
The techniques are available to tackle such problems, and the results
of the present study encourage us to believe that there is much more
to be learned from further studies of the out of equilibrium dynamics
of supersymmetric particle physics.

\acknowledgements
We thank Daniel Boyanovsky and Salman Habib for useful discussions.

%%%%%%%%%%%%%%%%%%%%%%%%%%%%%%%%%%%%%%%%%%%%%%%%%%%%%%%%%%%%%%%%

\begin{appendix}
\section{Renormalization: the Wess Zumino model}
\label{renormalization}
In order to renormalize the one-loop-equations we use the perturbation
scheme of \cite{Baacke:97a} which allows us to extract the divergences
from the leading orders. We recall some basic equations and refer the
interested reader to \cite{Baacke:98c,Baacke:97a} for further details.
We rewrite the mode equations (\ref{modesc}) and (\ref{fsec}) as
\begin{equation}\label{dgl}
\left[\frac{d^2}{dt^2}+\omega_{j0}^2\right]f_j(t)=-\calv_j(t) f_j(t)\kma
\end{equation}
with $j= A, B,\psi$ and
\begin{eqnarray} \label{defv1}
\calv_A(t)&=&\frac32 \lambda^2\left(\phi^2-\phi_0^2\right)
+3 m\lambda\left(\phi-\phi_0\right)=\calm_A^2(t)-m_{ A 0}^2 \kma \\
\label{defv2}
\calv_B(t)&=&\frac12 \lambda^2\left(\phi^2-\phi_0^2\right)
+ m\lambda\left(\phi-\phi_0\right)=\calm_B^2(t)-m_{ B 0}^2 \kma \\
\label{defv3}
\calv_\psi(t)&=& \lambda^2\left(\phi^2-\phi^2_0\right)+2\lambda
m\left(\phi-\phi_0\right)-i\lambda\dot \phi\nonumber\\
&=&\calm_\psi^2(t)-m_{\psi 0}^2-i\dot {\calm}_\psi(t)\pkt
\end{eqnarray}
With the initial conditions (\ref{init}) it is possible to
write the differential equation (\ref{dgl}) as an equivalent integral
equation
\begin{equation}
f_j(t)=e^{-i\omega_{j0}t}-\frac{1}{\omega_{j0}}\intt\sin\left[
\omega_{j0}(t-t')\right]\calv_j(t')f_j(t')\pkt
\end{equation}
We can now make a general ansatz for the mode functions via
\begin{equation}\label{hdef}
f_j(t)=e^{-i\omega_{j0}t}\left[1+h_j(t)\right] \pkt
\end{equation}
Inserting this ansatz into the integral equation one derives an
iteration that results in an expansion of the functions $h_j(t)$ in
orders of the potentials $\calv_j(t)$.

The perturbative expansion for the bosonic fluctuation integrals leads
to \cite{Baacke:97a} 
\bea \langle A^2\rangle &=& I_{-1}(m_{A0})-
\calv_{A}(t) I_{-3}(m_{A0})+\langle A^2\rangle_{\rm fin} \kma
\\
\langle B^2\rangle &=& I_{-1}(m_{B0})- \calv_{B}(t)
I_{-3}(m_{B0})+\langle B^2\rangle_{\rm fin} \pkt \eea The integrals
$I_{-k}, k=1,3$ are divergent. In dimensional regularization they
become \bea \label{defim3}
I_{-3}(m_{j0}^2)&=&\intk\frac{1}{4\omega^3_{j0}} \to
\frac{1}{16\pi^2}\left(\frac 2 \epsilon
  +\ln\frac{4\pi\mu^2}{m_{j0}^2}-\gamma\right) \kma \\ \label{defim1}
I_{-1}(m_{j0}^2)&=&\intk\frac{1}{2\omega_{j0}}\to
-m_{j0}^2I_{-3}(m_{j0}^2) -\frac{m_{j0}^2}{16\pi^2} \pkt \eea The
finite parts are defined as \bea \label{Aflfin} \langle
A^2\rangle_{\rm fin} &=&\intk\frac{1}{2\omega_{A0}}
\left\{|f_A(t)|^2-1+\frac{1}{2\omega^2_{A0}}\calv_A(t)\right\} \kma \\ 
\label{Bflfin} \langle B^2\rangle_{\rm fin}
&=&\intk\frac{1}{2\omega_{B0}}
\left\{|f_B(t)|^2-1+\frac{1}{2\omega^2_{B0}}\calv_B(t)\right\} \pkt
\eea The integration over the subtracted integrand is finite.  Note
that using the ansatz (\ref{hdef}) and working with the functions
$h_j(t)$ the leading divergence cancels explicitly, i.e., it is not
included from the outset.

For the fermions we obtain
\be \label{fermfl2}
\langle\bar\psi\psi\rangle=
I_{-3}(m^2)\left[2\ddot \calm_\psi+4\calm_\psi^3\right]+
\calf^\psi_{\rm fin} \kma
\ee
with
\bea \label{fermfl3}
\calf^\psi_{\rm fin}(t)&=&\frac{1}{16 \pi^2}
\left\{4m_{\psi0}^2\calm_\psi(t)-
\ln\left(\frac{m^2_{\psi0}}{m^2}\right)\left[4\calm_\psi^3(t)
+2\ddot \calm_\psi(t)\right]\right\}
+\langle \bar\psi\psi \rangle_{\rm fin} \pkt
\eea

It should be mentioned that the finite parts as defined here generally
become {\em singular} as $t \to 0$. These initial singularities can be
removed by a Bogoliubov transformation \cite{Baacke:98a}.  We do not
discuss this here, as it is not essential in the present context.

The total contribution of the fluctuations in the equation of motion
(\ref{oneloopeom}) has the form 
\bea \label{flucint}
\calf &=& \frac12 \lambda^2\phi\left[3\langle A^2\rangle
+\langle B^2\rangle\right] +
\frac{m\lambda}{2}\left[3\langle A^2\rangle+\langle B^2\rangle
+\frac 1 m\langle\bar\psi\psi\rangle\right] \pkt
\eea
Inserting the expressions for the expectation values given above we find
\begin{eqnarray}
\label{div}
{\cal F}&=&\lambda^2\left(
\ddot\phi-m^2-\frac{3}{2}\lambda m\phi^2-\frac12 \lambda^2\phi^3\right)
I_{-3}(m)\nonumber\\
&&+\Delta Z\ddot\phi+
\Delta {\rm F}+\Delta m\phi+\Delta\lambda\frac{3}{2}m\lambda\phi^2
+\frac12 \Delta\lambda\lambda^2\phi^3 + \calf_{\rm fin} \kma
\eea
where $\calf_{\rm fin}$ is defined by (\ref{flucint}), with all the 
expectation values replaced by their finite parts.  
Note that instead of the divergent integrals $I_{-k}(m_{i0})$ with the
different initial masses one has been  able to collect the divergent part into
one integral $I_{-3}(m)$ where $m$ is the renormalization mass, chosen here as
the common physical mass parameter of the model. 
The dependence on the initial conditions appears in the
finite terms
\begin{eqnarray}
\Delta Z&=&\frac{\lambda^2}{8\pi^2}\ln\frac{m_{\psi 0}^2}{m^2}\kma\\
\Delta {\rm F}&=&
-\frac{1}{32\pi^2}m\lambda
\left(3m_{ A 0}^2+m_{ B 0}^2-4m_{\psi 0}^2\right)\nonumber\\
&&-\frac{m^3\lambda}{32\pi^2}\left(
3\ln\frac{m_{ A 0}^2}{m^2}+\ln\frac{m_{ B 0}^2}{m^2}
-4\ln\frac{m_{\psi 0}^2}{m^2}\right)\kma\\
\Delta m&=&
-\frac{\lambda^2}{32\pi^2}
\left(3m_{ A 0}^2+m_{ B 0}^2-4m_{\psi 0}^2\right)\nonumber\\
&&-\frac{\lambda^2 m^2}{16\pi^2}\left(
6\ln\frac{m_{ A 0}^2}{m^2}+\ln\frac{m_{ B 0}^2}{m^2}
-6\ln\frac{m_{\psi 0}^2}{m^2}\right)\kma\\
\Delta\lambda&=&-\frac{\lambda^2}{32\pi^2}
\left(9\ln\frac{m_{ A 0}^2}{m^2}+\ln\frac{m_{ B 0}^2}{m^2}
-8\ln\frac{m_{\psi 0}^2}{m^2}\right)\pkt
\end{eqnarray}
The divergent part of (\ref{div}) can be written in the 
form
\begin{equation}
\calf_{\rm inf} \lambda^2\left[\ddot\phi-V'(\phi)\right]I_{-3}(m)
\pkt \end{equation}
It can be removed by adding to (\ref{WZlag}) a  counter term superlagrangian
\begin{equation} \label{WZconter}
{\cal L}_{\rm c.t.} = 
\left(\frac12 \delta Z S \cdot T S - \frac12 \delta m S \cdot S
- \frac13 \delta \lambda S \cdot S \cdot S \right)_F \; ,
\end{equation}
with
\begin{eqnarray}
\delta Z&=&-\lambda^2I_{-3}(m)\kma\\
\frac{\delta m}{m}&=&\frac{\lambda^2}{2}I_{-3}(m)\kma\\
\frac{\delta\lambda}{\lambda}&=&\frac{\lambda^2}{2}I_{-3}(m)\pkt
\end{eqnarray}
The equivalence between the mass and the coupling constant
counter term is a special feature of supersymmetry. With these counter terms
taken into account the equation of motion
becomes finite. Explicitly it is given by
\begin{eqnarray}
&&\left(1+\Delta Z\right)\ddot\phi+(m^2+\Delta m)\phi
+\frac{3}{2}m\lambda(1+\Delta\lambda)\phi^2\nonumber\\
&&+\frac12 \lambda^2(1+\Delta\lambda)\phi^3
+\Delta{\rm F}+{\cal F}_{\rm fin}=0\pkt
\end{eqnarray}

In the same way we have split ${\cal F}$ into divergent and finite parts
we can decompose the fluctuation part of the energy as
\begin{equation}
{\cal E}_{\rm fl}(t)={\cal E}_{\rm fl,div}(t)+{\cal E}_{\rm fl,fin}(t)\kma
\end{equation}
with
\begin{eqnarray}
{\cal E}_{\rm fl, div}&=&\intk\Biggl\{
\frac{1}{2}\left(\omega_{ A 0}+\omega_{ B 0}-2\omega_{\psi 0}\right)
\nonumber \\
&&+\frac 1 4\left[\frac{{\cal V}_A}{\omega_{ A 0}}
+\frac{{\cal V}_B}{\omega_{ B 0}}
-\frac{2}{\omega_{\psi 0}}\left(\calm_\psi^2(t)-m_{\psi 0}^2\right)\right]
\nonumber\\
&&\left.-\frac{1}{16}\left[\frac{{\cal V}_A^2}{\omega_{ A 0}^3}
+\frac{{\cal V}_B^2}{\omega_{ B 0}^3}-\frac{2}{\omega_{\psi 0}^3}
\left(\dot \calm_\psi^2(t)+\calm_\psi^4(t)+m_{\psi 0}^4
-2\calm_\psi^2(t)m_{\psi 0}^2
\right)\right]\right\} \kma\\
{\cal E}_{\rm fl, fin}&=&
\intk\frac{1}{2\omega_{A0}}\left[\frac 1 2|\dot f_A|^2
+\frac 1 2\left(k^2+\calm_A^2(t)\right)|f_A|^2-\omega_{A0}^2
-\frac{\calv_A}{2}+\frac{\calv_A^2}{8\omega_{A0}^2}\right]
\nonumber \\
&+&\intk\frac{1}{2\omega_{B0}}\left[\frac 1 2|\dot f_B|^2
+\frac 1 2\left(k^2+\calm_B^2(t)\right)|f_B|^2-\omega_{B0}^2
-\frac{\calv_B}{2}+\frac{\calv_B^2}{8\omega_{B0}^2}\right]\nonumber\\
&+&\intk\frac{1}{\omega_{\psi 0}}\left[i(\omega_{\psi 0}
-m_{\psi 0})(f_\psi\dot f_\psi^*-\dot f_\psi f_\psi^*)
-2\omega_{\psi 0}\calm_\psi (t)+2\omega_{\psi 0}^2+\calm_\psi^2(t) \right.\nonumber\\
&&\left.-m_{\psi 0}^2-\frac{1}{4\omega_{\psi 0}^2}\left(\dot \calm_\psi^2(t)
+\calm_\psi^4(t)+m_{\psi0}^4-2\calm_\psi^2(t)m_{\psi 0}^2\right)\right] \pkt
\end{eqnarray}
The energy has no initial singularity. In addition to the quadratic 
and logarithmic divergences we find here a quartic one. In dimensional
regularization it can be written as
\begin{equation}
I_1(m) = \intk\omega_{j 0}=-\frac{m_{j0}^4}{2}I_{-3}(m)-\frac{m_{j0}^4}{2}
\ln\frac{m_{j0}^2}{m^2}-\frac{3m_{j0}^4}{64\pi^2}\pkt
\label{defip1}
\end{equation}
In the same way as for the equation of motion we can now
evaluate the momentum integrals in ${\cal E}_{\rm div}$,
fix the counter terms, and find the finite contributions.
This leads to
\begin{eqnarray}
{\cal E}_{\rm div}&=&\lambda^2\left[\frac 1 2\dot\phi^2
-V(\phi)\right]I_{-3}(m)+\Delta {\rm F}+\Delta\Lambda\nonumber\\
&&+\frac 1 2 \Delta Z\dot\phi^2
+\frac 1 2 \Delta m\phi^2+\Delta\lambda\frac{1}{2}m\lambda\phi^3
+\frac 1 8\Delta\lambda\lambda^2\phi^4\kma
\end{eqnarray}
with
\begin{eqnarray}
\Delta\Lambda&=&\frac{1}{128\pi^2}(m_{ A 0}^4+m_{ B 0}^4
-2m_{\psi 0}^4).\nonumber\\
&&-\frac{m^4}{64\pi^2}\left(\ln\frac{m_{ A 0}^2}{m^2}
+\ln\frac{m_{ B 0}^2}{m^2}-2\ln\frac{m_{\psi 0}^2}{m^2}\right)\pkt
\end{eqnarray}
No divergent ``cosmological constant'' counter term is needed, 
as to be expected in a
supersymmetric model. After taking into account the counter term Lagrangian
(\ref{WZconter}) with the previously determined coefficients
the energy becomes finite and reads 
\begin{eqnarray}
{\cal E}_{\rm ren}&=&\frac{1}{2}(1+\Delta Z)\dot\phi^2
+\frac 1 2(m^2+\Delta m)\phi^2
+\frac{1}{2}m\lambda(1+\Delta\lambda)\phi^3\nonumber\\
&&+\frac{\lambda^2}{8}(1+\Delta\lambda)\phi^4+\Delta{\rm F}\phi
+\Delta\Lambda+{\cal E}_{\rm fl, fin}\pkt
\end{eqnarray}

\section{Renormalization of the large-$N$ model}

The expressions for the expectation values are analogous to those
of the simple Wess-Zumino model, and so is their perturbative expansion.
Of course now the fluctuation masses $\calm_j(t)$, given by
Eqs. (\ref{lnflmass1}),(\ref{lnflmass2}) and (\ref{m-eqn}), depend on the
expectation values of the fluctuations, and so do the potentials
$\calv_j(t)$, defined in analogy to (\ref{defv1})-(\ref{defv3}).

In  the following we will need the sum and the difference of the
bosonic fluctuations.  For the sum we obtain
\be
\langle A^2+B^2\rangle=
-2I_{-3}(m^2)\left(m+\lambda \phi\right)^2 + \calf_{\rm fin}^+(t)
\kma
\ee
with
\bea \nonumber
\calf_{\rm fin}^+(t)&=& \frac{1}{16 \pi^2}
\left\{
\ln\left(\frac{m^2_{A0}}{m^2}\right)\calm^2_A(t)
+\ln\left(\frac{m^2_{B0}}{m^2}\right)\calm^2_B(t)
-m_{A0}^2-m_{B0}^2\right\}  
\\ \label{deffplus}
&&+\langle A^2\rangle_{\rm fin}+\langle B^2\rangle_{\rm fin}
\pkt \eea
$\langle A^2\rangle_{\rm fin}$ and $\langle B^2\rangle_{\rm fin}$ 
are again defined by (\ref{Aflfin}) and (\ref{Bflfin}).
For the difference we find
\be
\langle A^2-B^2\rangle
=
-2I_{-3}(m^2)\left[M\lambda\phi+
\frac{1}{2}\kappa\lambda \phi^2
+\frac{1}{2}\lambda^2\langle A^2-B^2\rangle \right]
 + \tilde\calf^-_{\rm fin}(t) \kma
\ee
with
\bea  \nonumber
\tilde\calf_{\rm fin}^-(t)&=& \frac{1}{16 \pi^2}
\left\{
\ln\left(\frac{m^2_{A0}}{m^2}\right)\calm^2_A(t)
-\ln\left(\frac{m^2_{B0}}{m^2}\right)\calm^2_B(t)
-m_{A0}^2+m_{B0}^2\right\}  
\\ 
&&+\langle A^2\rangle_{\rm fin}-\langle B^2\rangle_{\rm fin}
\pkt
\eea
Obviously the equation for $\langle A^2-B^2\rangle$ 
is implicit, as to be expected in the large-$N$
framework. The logarithms of the ratios $m^2_{j0}/m^2$ arise
from replacing the masses $m_{j0}$ by the common renormalized mass
$m$ of all the component fields in the integral $I_{-3}(m^2)$. 
It is convenient to rewrite these equations in terms of
$\calm_-^2$, as introduced in Eq. \eqn{m-eqn}.
We get
\be
\langle A^2-B^2\rangle
=
-2\left[I_{-3}(m^2)-\frac{1}{16\pi^2}
\ln\left(\frac{m_{A0}m_{B0}}{m^2}\right)\right]
\calm_-^2
 + \calf^-_{\rm fin}(t) \kma
\label{Fminusexp}
\ee
with
\bea \nonumber
\calf_{\rm fin}^-(t)&=& -\frac{1}{16 \pi^2}
\ln\left(\frac{m^2_{A0}}{m_{B0}^2}\right)\calm_\psi^2(t)
-\frac{1}{16 \pi^2}\left(m_{A0}^2-m_{B0}^2\right)  
\\ 
&&+\langle A^2\rangle_{\rm fin}-\langle B^2\rangle_{\rm fin}
\pkt
\eea
The expectation value of the fermionic fluctuations
again decomposes as (\ref{fermfl2}) and (\ref{fermfl3}).

We now introduce multiplicative renormalization factors by replacing
\bea
\lambda \to Z_\lambda \lambda
\\
A \to Z_A A
\\
\phi \to Z_\phi \phi
\eea
and similarly for all other quantities.
We first note that there is no divergent term that would require an
infinite renormalization of the mass $m$. This mass occurs in the 
combination
$m+\lambda\phi$ which now gets replaced by
$m+Z_\lambda Z_\phi\lambda\phi$. We conclude that also the second
term stays unrenormalized, and so
\be \label{Zlamphi}
Z_\lambda Z_\phi=1
\pkt
\ee 
We next analyze the mass $\calm_-^2(t)$.
We have
\be \label{calmmbare}
\calm_-^2(t)=Z_MZ_\lambda Z_\phi M\lambda\phi+
\frac{1}{2}Z_\kappa Z_\lambda Z_\phi^2\kappa\lambda\phi^2
+\frac{1}{2} Z_A^2 Z_\lambda^2 \lambda^2
\langle A^2-B^2 \rangle \pkt
\ee
We have assumed $Z_A=Z_B$. Further, from (\ref{Fminusexp}), we have
\bea
&&\calm_-^2\left[1+\lambda^2Z_A^2Z_\lambda^2
I_{-3}(m^2)\right]=
Z_MZ_\lambda Z_\phi M\lambda\phi+
\frac{1}{2}Z_\kappa Z_\lambda Z_\phi^2\kappa\lambda\phi^2
\\ \nonumber &&
+\lambda^2Z_A^2Z_\lambda^2\frac{1}{16\pi^2}
\ln\left(\frac{m_{A0}m_{B0}}{m^2}\right)
\calm_-^2
+ \frac{1}{2} Z_A^2 Z_\lambda^2 \lambda^2\calf^-_{\rm fin}
\pkt
\eea
In order to get analogous relations for the finite
quantities as for the unrenormalized ones we require
\be
\left[1+\lambda^2Z_A^2Z_\lambda^2
I_{-3}(m^2)\right]=Z_A^2Z_\lambda^2=Z_M=Z_\kappa Z_\phi
\kma\ee
having used Eq. \eqn{Zlamphi}. The first relation yields
\be \label{ZAlambdasq}
Z_A^2Z_\lambda^2=\frac{1}{1-\lambda^2I_{-3}(m^2)} \pkt
\ee
We then have further
\be
Z_M=Z_\kappa Z_\phi =\frac{1}{1-\lambda^2I_{-3}(m^2)}
\pkt
\ee
The renormalized equation for the mass $\calm_-$
becomes
\be 
 \calm_-^2(t)\left[1+\frac{\lambda^2}{16\pi^2}
\ln\left(\frac{m_{A0}m_{B0}}{m^2}\right)\right]
=M\lambda\phi(t)+\frac{1}{2}\kappa\lambda \phi^2(t)
+\frac{1}{2}\lambda^2 \calf^-_{\rm fin} \kma
\ee
or
\be \label{MAmBren}
 \calm_-^2(t)
=\calc\left[M\lambda\phi(t)+\frac{1}{2}\kappa\lambda \phi^2(t)
+\frac{1}{2}\lambda^2 \calf^-_{\rm fin}\right] \kma
\ee
with
\be
\calc=\frac{1}{\displaystyle 1+\frac{\lambda^2}{16\pi^2}
\ln\left(\frac{m_{A0}m_{B0}}{m^2}\right)} \pkt
\ee
Using Eq. \eqn{Zlamphi} we also have the renormalized relations
\bea \label{renmasses}
\calm_{A}^2(t) &=& \left[m+\lambda\phi(t)\right]^2 +
\calm_-^2(t) \kma
\\
\calm_{B}^2(t) &=& \left[m+\lambda\phi(t)\right]^2 -
\calm_-^2(t) \kma
\\
\calm_\psi(t)&=&m+\lambda\phi(t)
\pkt\eea

We now turn to the equation of motion for the mean field
$\phi(t)$, Eq. \eqn{mean}.
Introducing the renormalization factors we obtain
\bea \label{eqphirenbare}
&&Z_\phi \ddot\phi+\frac{Z_M}{Z_\lambda\lambda}(M+\kappa\phi)
\calm_-^2+Z_\lambda^2Z_\phi Z_\psi^2
\lambda^2I_{-3}(m^2)\ddot\phi
\\
&&+Z_\lambda Z_A^2\lambda\left(m+\lambda\phi\right)
\langle A^2+B^2 \rangle
+Z_\lambda Z_\psi^2\frac{\lambda}{2}\langle\bar\Psi\Psi\rangle =0
\pkt\eea
We first consider the last two terms.
These terms are quadratically divergent. This ``tadpole'' contribution
necessitates a cancellation between bosons and fermions. This works only if
we postulate
\be
Z_\psi=Z_A=Z_B
\kma\ee
a relation indeed required by supersymmetry.
Then
\be
Z_\lambda Z_A^2\lambda\left(m+\lambda\phi\right)
\langle A^2+B^2 \rangle
+Z_\lambda Z_\psi^2\frac{\lambda}{2}\langle\bar\Psi\Psi\rangle =
Z_\lambda Z_A^2\lambda\left[\calm_\psi\langle A^2+B^2 \rangle+
\frac{1}{2}\langle\bar\Psi\Psi\rangle\right] \pkt
\ee
Inserting the decomposition into divergent parts and fluctuation 
integrals the expression in brackets yields
\be
[\dots]=I_{-3}(m^2)\ddot \calm_\psi+
(m+\lambda \phi)\calf^+_{\rm fin}+
\frac{1}{2}\calf^f_{\rm fin}
\kma\ee
so that the equation of motion for $\phi$ takes the form
\bea \label{eqphirenbrut}
&&Z_\phi \ddot\phi+\frac{Z_M}{Z_\lambda\lambda}(M+\kappa\phi)
\calm_-^2+Z_\lambda^2Z_\phi Z_A^2
\lambda^2I_{-3}(m^2)\ddot\phi
\\
&&+
\frac{1}{2}Z_\lambda Z_A^2\lambda\calf^f_{\rm fin}
+Z_\lambda Z_A^2 \lambda(m+\lambda\phi)\calf^+_{\rm fin}=0
\pkt
\eea
The coefficients of the two terms 
containing $\ddot\phi$ combine into
\be
Z_\phi\left[1+Z_\lambda^2 Z_\psi^2
\lambda^2I_{-3}(m^2)\right]=Z_\phi
\frac{1}{1-\lambda^2I_{-3}(m^2)} \pkt
\ee
Here we have used Eq. \eqn{ZAlambdasq}.
The factor of the $\ddot\phi$ term agrees with that of the second term
in Eq. \eqn{eqphirenbrut}
\be
\frac{Z_M}{Z_\lambda}=Z_M Z_\phi=Z_\phi
\frac{1}{1-\lambda^2I_{-3}(m^2)}
\pkt
\ee
Likewise for the fourth term in  Eq. \eqn{eqphirenbrut}
we obtain
\be
Z_\lambda Z_A^2=Z_\phi Z_\lambda^2 Z_A^2=Z_\phi
\frac{1}{1-\lambda^2I_{-3}(m^2)}
\pkt\ee
The same factor is obtained for the fifth term, so the 
renormalized equation reads
\be \label{eqphiren}
\ddot\phi+\frac{1}{\lambda}(M+\kappa\phi)
\calm_-^2
+
\frac{1}{2}\lambda\calf^f_{\rm fin}
+ \lambda(m+\lambda\phi)\calf^+_{\rm fin}=0
\pkt\ee

The various factors $Z_j$ are not yet completely determined.
This is to be expected, since the equations of motion do not
depend on the absolute normalization of the Lagrangian.
When including the $Z$ factors and using the various relations derived
so far, the energy density is given by
\bea \nonumber
\cale&=&\frac12 Z_\phi^2\dot\phi^2+ \frac12 Z_M^2 Z_\phi^2\left(M\phi
+\frac{\kappa}{2}\phi^2\right)^2
\\ \label{ebare} &&
+\frac12 Z_A^2\langle \dot A^2+k^2A^2+\calm_A^2(t) A^2\rangle
+\frac12 Z_A^2\langle \dot B^2+k^2B^2+\calm_B^2(t) B^2\rangle
\\ &&\nonumber -Z_\lambda^2Z_A^4
\frac{\lambda^2}{8}\left(\langle A^2-B^2\rangle\right)^2
+Z_A^2\frac{1}{2}\left\langle\bar\psi\left(-i\vec\gamma \cdot \vec\nabla
+\calm_\psi(t)\right)\psi\right\rangle
\pkt
\eea
The bosonic fluctuation energies defined as as
\bea
\label{energya}
\cale_A&=&
\frac12 \langle \dot A^2+k^2A^2+\calm_A^2(t) A^2\rangle 
\\\nonumber
&=&\intk\frac{1}{2\omega_{A0}}\left\{\frac{1}{2}|\dot f_A|^2+
\frac{1}{2}\left[k^2+\calm_A^2(t)\right]|f_A|^2\right\} \kma
\eea
and analogously for the field $B$.
Its expansion in terms of divergent integrals
reads
\bea
\cale_A&=&\frac 1 2 I_1(m_{A0}^2)+\frac{1}{2}I_{-1}(m_{A0}^2)
\calv_A(t)-\frac{1}{4}I_{-3}(m_{A0}^2)\calv_A^2(t)+\cale_{A,\rm fin}
\\ \nonumber
&=&
-\frac{1}{4}I_{-3}(m^2)\calm_A^4+
\frac{1}{128\pi^2}m_{A0}^4
-\frac{1}{32\pi^2}m_{A0}^2\calm_A^2
-\frac{1}{64\pi^2}\ln\left(\frac{m^2}{m_{A0}^2}\right)\calm_A^4
+\cale_{A,\rm fin}
\pkt
\eea
Here we have used Eqs.~(\ref{defim3}), (\ref{defim1}), and (\ref{defip1}).
The finite part is defined by subtracting the quantity
\be
\cale_{A,\rm div} = \intk \frac{1}{2\omega_{A0}} 
\left\{\omega_{A0}^2+\frac{1}{2}\calv_A-\frac{1}{8\omega_{A0}^2} \calv_A^2 \right\} \kma\ee
from the expression (\ref{energya}).

The fermionic fluctuation energy is defined as
\bea  \label{eflfermdef}
&&\cale_\psi=\frac{1}{2}\left\langle\bar\psi\left(-i\vec\gamma 
\cdot \vec\nabla +\calm_\psi(t)\right)\psi\right\rangle
\\ \nonumber
&&=
\intk\frac{1}{2\omega_{\psi 0}}\left\{i\left[\omega_{\psi0}-m_{\psi0}\right]
\left(f(k,t)\dot f^*(k,t)-\dot f(k,t) f^*(k,t)\right)
-2\omega_{\psi0}\calm_\psi(t)\right\}
\pkt\eea
It is expanded as
\bea
\cale_\psi&=&\frac{1}{2}\left[\dot\calm_\psi^2+\calm_\psi^4\right]
\left[I_{-3}(m^2)+\frac{1}{16\pi^2}\ln\left(\frac{m^2}{m_{\psi0}^2}\right) 
\right] \nonumber
\\  &&-\frac{m_{\psi0}^4}{64\pi^2}+\frac{m_{\psi0}^2}{16\pi^2}\calm_\psi^2
+\cale_{\psi,\rm fin} \kma
\eea
where $\cale_{\psi,\rm fin}$ is defined by subtracting
\be
\cale_{\psi,\rm div} = \intk \frac{1}{2\omega_{\psi 0}} \left\{
-2\omega_{\psi 0}^2-(\calm_\psi^2-m_{\psi 0}^2)+\frac{\dot\calm_\psi^2}{4\omega_{\psi 0}^2}
+\frac{(\calm_\psi^2-m_{\psi 0}^2)^2}{4\omega_{\psi 0^2}} \right\} \kma
\ee
from the integral of Eq.~(\ref{eflfermdef}).

With these preparations we are ready to discuss the divergences of the
energy density.

The term proportional to $\dot\calm_\psi^2=\lambda^2\dot\phi^2$ 
adds to the kinetic
term $Z_\phi^2\dot\phi^2$. The total factor (up to a finite term, see
below) is $Z_\phi^2+Z_A^2\lambda^2 I_{-3}$. If we compare to the
relations for the $Z_j$ found in the previous section we find that
the prefactor of $\dot\phi^2$ becomes finite if we choose
$Z_A=1$ and as a consequence $Z_\phi^2=1-\lambda^2I_{-3}$. 

The second and the fifth term in the general expression
for the energy, Eq. \eqn{ebare}
can be rewritten, using Eq. \eqn{calmmbare} as
\bea
&&  \frac12 Z_M^2 Z_\phi^2\left(M\phi
+\frac{\kappa}{2}\phi^2\right)^2- Z_\lambda^2Z_A^4
\frac{\lambda^2}{8}\left(\langle A^2-B^2\rangle\right)^2
\\ \nonumber
&=& \frac12 \left[Z_MZ_\phi\left(M\phi
+\frac{\kappa}{2}\phi^2\right)-\frac{\lambda}{2}
Z_\lambda Z_A^2\langle A^2-B^2\rangle\right]\frac{1}{Z_\lambda\lambda}
\calm_-^2
\\ \nonumber
&=& Z_MZ_\phi\left(M\phi
+\frac{\kappa}{2}\phi^2\right)\frac{1}{Z_\lambda\lambda}
\calm_-^2-\frac{1}{2Z^2_\lambda\lambda^2}
\calm_-^4 \pkt
\eea
We note that $Z_MZ_\phi/Z_\lambda=1$ for the choice
\eqn{choice22}, so that the first term on the right hand
side is finite. The second term $-\calm_-^4/2Z_\lambda^2\lambda^2$
has a factor
$Z_\lambda^{-2}=1-\lambda^2 I_{-3}$.
This combines with the infinite parts of $\cale_A,\cale_B$
and $\cale_\psi$ as
\bea 
-\frac12 \calm_-^4\left[\frac{1}{\lambda^2}-I_{-3}(m^2)\right]
-\frac{1}{4}\left(\calm_\psi^2+\calm_-^2\right)^2I_{-3}(m^2)&&
\\ \nonumber
-\frac{1}{4}\left(\calm_\psi^2-\calm_-^2\right)^2I_{-3}(m^2)
+\frac{1}{2}\calm_\psi^4I_{-3}(m^2)&=&
-\frac{\calm_-^4}{2\lambda^2}
\pkt\eea
So all infinite terms have cancelled, and the choice of renormalization
constants
\bea \label{choice21}
Z_A&=&Z_B=Z_\psi=1
\kma \\
\label{choice22}
Z_\lambda^2&=&Z_\phi^{-2}=Z_M=\frac{1}{1-\lambda^2I_{-3}(m^2)}
\kma \\
\label{choice23}
Z_\kappa&=&\frac{1}{\left[1-\lambda^2I_{-3}(m^2)
\right]^{3/2}} \kma
\eea
based on the analysis of the equations of motion and of the energy
momentum tensor. It remains to collect the
finite terms. We find
\bea \nonumber
\cale&=&\frac12 \dot \phi^2\left[1-\frac{\lambda^2}{32\pi^2}
\ln\left(\frac{m_{\psi 0}^2}{m^2}\right)\right] 
+\left(M\phi+\frac{\kappa}{2}\phi^2\right)
\frac{\calm_-^2}{\lambda} -\frac{1}{2\lambda^2}\calm_-^4
\\&&
+\tilde\cale_{A,\rm fin}+\tilde\cale_{B,\rm fin}+
\tilde\cale_{\psi,\rm fin} \kma
\eea
with
\bea
\tilde \cale_{A,\rm fin}
&=&\cale_{A,\rm fin}+\frac{1}{128\pi^2}\left[
m_{A0}^4-4m_{A0}^2\calm_A^2+2\ln\left(\frac{m_{A0}^2}{m^2}\right)
\calm_A^4\right]
\\ 
\tilde \cale_{B,\rm fin}
&=&\cale_{B,\rm fin}+\frac{1}{128\pi^2}\left[
m_{B0}^4-4m_{B0}^2\calm_B^2+2\ln\left(\frac{m_{B0}^2}{m^2}\right)
\calm_B^4\right]
\\ 
\tilde \cale_{\psi,\rm fin}
&=&\cale_{\psi,\rm fin}-\frac{1}{64\pi^2}\left[
m_{\psi 0}^4-4m_{\psi 0}^2\calm_\psi^2
+2\ln\left(\frac{m_{\psi 0}^2}{m^2}\right)
\calm_\psi^4\right]
\pkt
\eea
We note again that no `cosmological constant' counter term has 
to be introduced.
\end{appendix}

\end{document}